\documentclass[journal]{IEEEtran}
\usepackage{mathrsfs}
\usepackage{subfigure}
\usepackage{algorithm}
\usepackage{algorithmic}
\usepackage{bm}
\usepackage{amsmath}
\usepackage{amssymb}
\usepackage{graphics}
\usepackage{graphicx}
\usepackage{epstopdf}
\usepackage{url}
\usepackage{cite}
\usepackage{diagbox}
\usepackage{multirow}
\makeatletter
\@addtoreset{equation}{section}
\makeatother
\renewcommand{\theequation}{\arabic{section}.\arabic{equation}}
\ifCLASSINFOpdf
\else
\fi
\begin{document}
%\bibliographystyle{plain}
\title{Blockchain Systems, Technologies and Applications: A Methodology Perspective}

\author{Bin~Cao,~\IEEEmembership{Member,~IEEE},
        Zixin~Wang,
        Long~Zhang,
        Daquan~Feng,
        Mugen~Peng,~\IEEEmembership{Fellow,~IEEE},
        and~Lei~Zhang,~\IEEEmembership{Senior~Member,~IEEE}% <-this % stops a space
\thanks{ Bin Cao, Zixin Wang, and Mugen Peng are with State Key Laboratory
of Networking and Switching Technology, Beijing University of Posts
and Telecommunications, Beijing, 100876, China. E-mail: caobin@bupt.edu.cn, wzxph@foxmail.com, pmg@bupt.edu.cn.}% <-this % stops a space
\thanks{Long  Zhang is with the National Key Laboratory of Science
and Technology on Communications, University of Electronic Science and
Technology of China, Chengdu 611731, China. E-mail: zhanglong3211@yeah.net}% <-this % stops a space
\thanks{Daquan Feng is with the Guangdong Province Engineering Laboratory
for Digital Creative Technology and Guangdong Key Laboratory of Intelligent Information Processing, Shenzhen University, Shenzhen 518060, China.
E-mail: fdquan@szu.edu.cn}
\thanks{Lei Zhang is with the James Watt School of Engineering, University of
Glasgow, Glasgow G12 8QQ, U.K. E-mail: lei.zhang@glasgow.ac.uk}
}


\markboth{Journal of \LaTeX\ Class Files,~Vol.~14, No.~8, August~2015}%
{Shell \MakeLowercase{\textit{et al.}}: Bare Demo of IEEEtran.cls for IEEE Journals}



\maketitle

\begin{abstract}
 In the past decade, blockchain has shown a promising vision greatly to build the trust without any powerful third party in a secure, decentralized and salable manner. However, due to the wide application and future development from cryptocurrency to Internet of Things, blockchain is an extremely complex system enabling integration with mathematics, finance, computer science, communication and network engineering, etc. As a result, it is a challenge for engineer, expert and researcher to fully understand the blockchain process in a systematic view from top to down. First, this article introduces how blockchain works, the research activity and challenge, and illustrates the roadmap involving the classic methodology with typical blockchain use cases and topics. Second, in blockchain system, how to adopt stochastic process, game theory, optimization, machine learning and cryptography to study blockchain running process and design blockchain protocol/algorithm are discussed in details. Moreover, the advantage and limitation using these methods are also summarized as the guide of future work to further considered. Finally, some remaining problems from technical, commercial and political views are discussed as the open issues. The main findings of this article will provide an overview in a methodology perspective to study theoretical model for blockchain fundamentals understanding, design network service for blockchain-based mechanisms and algorithms, as well as apply blockchain for Internet of Things, etc.
\end{abstract}


% \fi
\IEEEpeerreviewmaketitle

\section{Introduction}
\IEEEPARstart{O}{riginally }  proposed as the backbone technology
of Bitcoin \cite{chapter1-1}, Ethereum \cite{chapter1-2}, and many other digital currencies \cite{chapter1-3}, blockchain has become a revolutionary
decentralized data management framework that establishes consensuses and agreements in a trust-less and distributed environment  \cite{7163223}.
%As the underlying technology of Bitcoin \cite{chapter1-1}, Ethereum\cite{chapter1-2} and other cryptocurrencies \cite{chapter1-3} \cite{Tschorsch2016}, blockchain \cite{7163223} has become a promising decentralized solution for trust and security problems.
In addition to the soaring in the finance sector, blockchain has been attracted much attention from many other major industrial sectors  ranging from supply chain \cite{8674550}, transportation \cite{transportation7795984}, entertainment \cite{retail8096051}, retail \cite{8640264}, healthcare \cite{2005.10103}, information management \cite{8658586} to financial services \cite{SANGWAN2020265}, etc. As such,  Gartner forecasts that by 2030, blockchain will generate an annual business value of more than US \$3 trillion, and envisions that 10\% to 20\% of global economic infrastructure will be running on blockchain-based systems \cite{Gartner2017}.
%Recently, blockchain has been considered as a feasible solution for addressing the challenges of trusty and security in the fields of Internet of Things (IoT) \cite{8731639}, Mobile Edge Computing (MEC) \cite{8998330} and supply chain \cite{8674550}.

%Essentially, blockchain is a decentralized ledge management system for recording and validating data. Without involvement of an authority or third party, all peer nodes work together to maintain public ledge with aim of realizing trust, security, transparency and immutability.
%To enable these expected functions, an technology is needed to integrate various technologies, such as Peer-to-Peer (P2P) transmission protocols [references], consensus mechanisms \cite{SALIMITARI2020100212}, and cryptographic algorithms \cite{chapter7Introduction8} .

Fundamentally, blockchain is a decentralized ledge management system for recording and validating transaction. It allows two parties to complete a  transaction in a peer-to-peer (P2P) network \cite{8428402}.  Without involvement of an authority or third party, all peer nodes work together to maintain public ledge with aim of realizing trust, security, transparency and immutability.  The recorded transaction in blockcahin can be any form of data which involves the ownership  transfer or sharing  of  resource,
where it can be tangible such as money, houses, cars, land, or intangible   copyright, digital documents, and intellectual property, etc.

Essentially, blockchain is built on a physical  network that relies on the communications, computing and caching, which serves the basis of blockchain functions such as incentive mechanism or consensus. As such,  blockchain systems can be depicted as a two-tier architecture: an infrastructure layer and a blockchain layer.
The infrastructure layer is the underlying entity and responsible for maintaining P2P network, building connection through wired/wireless communication, computing and storing data. The top is the blockchain layer that can realize trust and security functions based on underlying information exchanging.
More specifically,  blockchain features several key components which are summarized as: transaction, block and chain of blocks. \emph{Transaction} contains the information requested by the client and need be recorded by public ledge; \emph{block} securely records an amount of transactions or other useful information; using consensus mechanism, blocks are linked orderly to constitute a \emph{chain of blocks}, which indicates logical relation among the blocks to construct blockchain. As a core function of the blockchain, the consensus mechanism works in the blockchain layer   ensures a clear sequence of transactions and ensures the integrity and consistency of the blockchain across geographically distributed nodes \cite{2003.13083}.
State of a blockchain is updated when a valid transaction  is recorded on chain, and smart contracts\footnote{The smart contract is a computer program designed to digitally facilitate direct negotiations or contract terms between users when certain conditions are met \cite{8758979}. }  can be used to automatically trigger transactions under certain conditions \cite{1810.04699}.  Therefore, due to its autonomy and efficiency, smart contracts are being used for a wide range of purposes, from self-managed identities on public blockchains to allowing automated business collaboration on blockchains.

%As an open distributed ledger, the blockchain information can be audited by all users, which can effectively improve network security and privacy.

Driven by the continuous development of 5G technology, more and more services have been launched to improve network performance and user experience. Importantly, features such as data immutability and transparency are the key factors to ensure the successful launch of new services such as IoT data collection, driverless cars, drones, and federal learning.  Blockchain is   regarded as the most promising to meet these new requirements with its decentralization, openness, tamper resistance, anonymity and traceability. Therefore, in order to more thoroughly explore the potential of blockchain and make it better serve the requirements of modern networks, it is necessary to comprehensively and systematically understand blockchain from top to bottom. Methodology advocates going to the bottom of the problem, digging into the essence behind the phenomenon, and forming a theoretical system with a certain depth. The inherent contradictions of the problem can be revealed and the fundamental solution can be found from methodology perspective.  Therefore,  methodology can be well suited to the research of blockchain system performance to reveal the principles and problems of blockchain running process and blockchain protocol/algorithm design in the blockchain system, and provide theoretical support for solving specific problems. Consequently, this paper outlines the theoretical model research of blockchain basic knowledge, the design of network services based on blockchain mechanisms and algorithms, and the deployment of blockchain-based applications in practical systems from a methodological perspective, as well as further outlines the application of methodology in blockchain from multiple dimensions such as advantages, limitations, case studies and challenges. It aims to provide a comprehensive and clear overview for researchers in the blockchain field.

%To reach consistency, consensus mechanisms are designed to ensure consistent state among data recording. Meanwhile, incentive mechanisms are needed to motivate nodes to contribute their efforts to reach consistency \cite{8863721}. Besides, as a feasible way to create and execute complicated transactions, smart contracts are expected to automatically implement decisions for the specific application based on predefined rules [references].
\subsection{How Does Blockchain Work}
%In order to illustrate the blockchain work more clearly, the basic components: transaction, block and chain of blocks will be illustrated. The generated transaction can be seen as data that need be recorded by public ledge; block records an amount of transactions and useful information for security; using consensus mechanism, blocks are linked orderly to constitute a chain of blocks, which indicates logical relation among the blocks to construct blockchain.

%For a blockchain system, the infrastructure layer and blockchain layer are interrelated and interact on each other cooperatively. The infrastructure layer is designed to realize the communication and interconnection of human, things and data, so as to help blockchain layer reach consensus and other blockchain-based functions. Using consensus mechanism, cryptographic algorithm, etc, the blockchain layer aims to address trust, security and scalability issues in a decentralization manner.

\begin{figure*}
 % \captionsetup{font={footnotesize}}
  \centering
  \includegraphics[width=18cm]{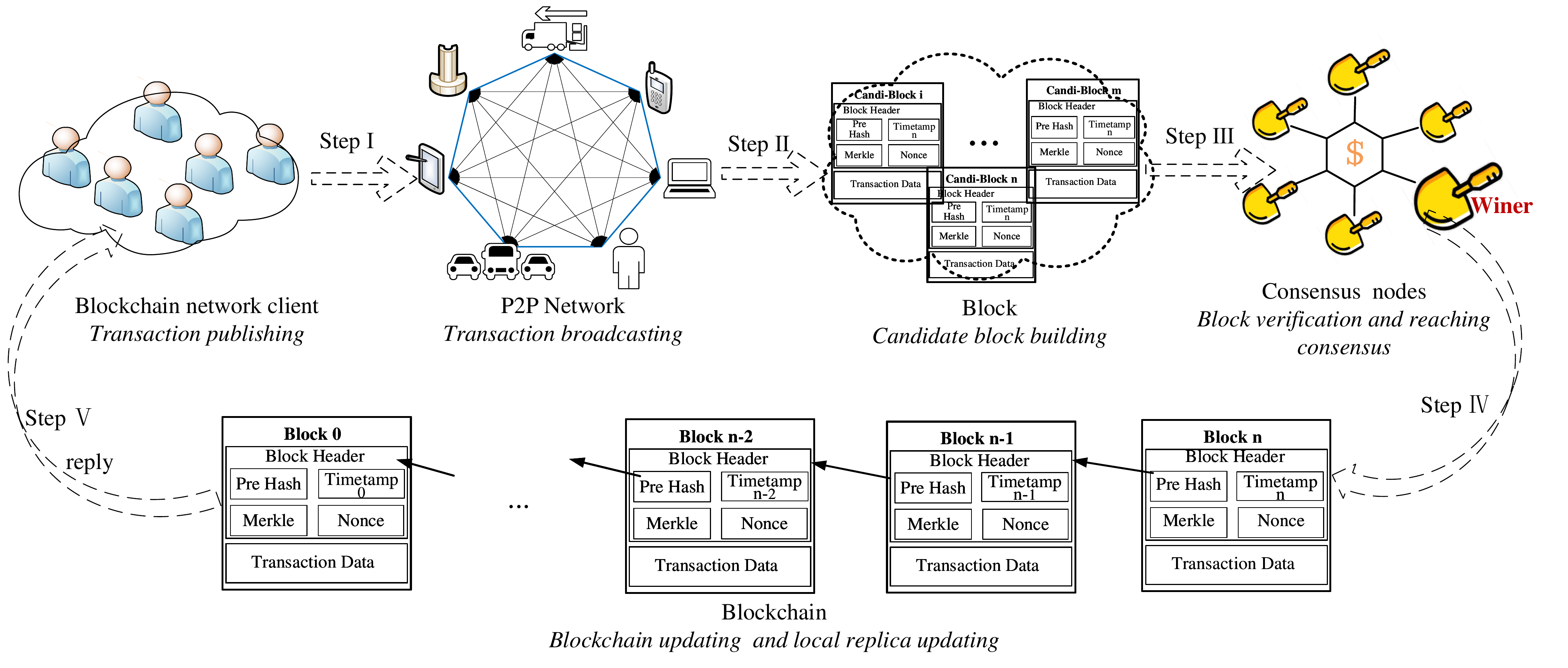}
  \caption{An overview of blockchain workflow}
  \label{fig:overview of blockchain workflow}
\end{figure*}

The infrastructure layer and blockchain layer are interrelated and interact on each other cooperatively,  while the detailed procedures can vary among different blockchain systems. However, for all blockchain systems, they have to follow the following basic steps, as   shown in Fig. 1.
%once one node connects P2P network, it should know the network information to verify the validity of incoming transactions. Therefore,
Firstly, blockchain clients generate transactions and broadcast them to the P2P network in Step I. More than one nodes may bundle different subset of unverified transactions into their candidate blocks  in Step II. Afterwards, all nodes perform incentive and consensus mechanisms to find the winner whose candidate block would be announced as a new block in Step III. Then, the new block  will be  inserted into all node's local ledgers in Step IV.
%timestamped after being authenticated by the blockchain network and are then included in a block encrypted by a hash process that links to the hash of the previous block and merged into it in Step IV.
This mechanism connects multiple blocks together and builds a chronological chain. In particular, the process of hashing a new block always contains metadata about the hash value of the previous block, which makes the linked data highly non-modifiable. At last, after the transaction is stored in the blockchain, the client can request the consensus network to confirm whether a transaction is in the blockchain.

Through this workflow, the transaction is finally agreed by the majority of nodes and recorded in blockchain, where malicious nodes cannot subvert the consensus results. This decentralized architecture ensures robust and safe operations on the blockchain, with the advantages of tamper resistance and no single point of failure. It can be seen from the Fig. 1 that each  blockchain node in the blockchain system undertakes all or part of functions, such as communication between nodes, maintenance of P2P network, and computation of consensus mechanism.
Thus, the underlying communication, network and computation is crucial to establish effective and secure blockchain system. This encourages us to study how communication, networking and computing affect blockchain systems. Fortunately, some classic methodologies can provide good ideas, such as stochastic  process for block generation and nodes communication, machine learning  for P2P network performance improving, and optimization theory for resource allocation. Therefore, it is feasible and valuable to explore the interaction process of communication, network and computing and their impact on the blockchain system from the perspective of methodology, and even can provides help for revealing the essential problems in the operation process of the blockchain system.

\subsection{Existing Surveys}
%As a technology with great potential to build a distributed, transparent, and tamper-proof network, blockchain is a research hotspot in industry and academia.
Recognising the wide applications of blockchain technology, a novel survey paper can help researchers in various fields to build good foundations on the subject to guide actual developments.
Recently,  several   work have reviewed the advanced development of blockchain from various views.

For security and privacy, T. Salman et al. in \cite{8428402} present  blockchain-based security services in authentication, confidentiality, privacy and access control, etc. N. Waheed et al. in  \cite{2002.03488} summarize  the research efforts of using machine learning algorithm and blockchain technology to address security and privacy problems in the field of Internet of Things in the past few years.  M. Conti et al. in \cite{8369416} focus  on the security and privacy threats of Bitcoin, and discusses the feasibility and limitations of potential solutions.   M. Saad et al. in
\cite{9019870}
%\cite{Saad2020}
focus  on how attacks effect public blockchain and discusses the relationships between a sequence of possible attacks.

%By enabling the integration of blockchain and other advanced technologies, some of work explore potential applications and research challenges in IoT \cite{Ferrag2019,platformsforIoTuse-cases}, smart city \cite{Xie2019}, cloud computing \cite{Gai2020}, edge computing \cite{Yang2019}, and fog computing\cite{9104970}.

Besides, as the core of blockchain, consensus determines the performance and security of the blockchain in many ways,  M. S. Ferdous et al. in \cite{ConsensusAlgorithms} utilize  comprehensive taxonomy of properties to analyze a wide range of consensus algorithms, and examines in detail the meaning of the different problems that are still prevalent in the consensus algorithm. W. Wang et al. in \cite{Wang2019} review  the state of the art consensus protocols and game theory in mining strategy management.

In the context of artificial intelligence (AI),  Y. Liu et al. in \cite{Liu2020} discuss  feasible solutions integrating blockchain and machine learning for communications and networking system. Y. Xiao et al. in \cite{Xiao2020} introduce  the classic theory of fault tolerance and analyzes blockchain consensus protocols using a five-component framework.

For the scalability of the blockchain,  J. Xie et al. in \cite{8823874} study the scalability of the blockchain system, analyzes the scalability from the perspective of throughput, storage and network, and introduces the existing enabling technology of the scalable blockchain system.  H. T. M. Gamage et al. in \cite{TechnologyConcepts} discuss  issues of the existing blockchains such as 51\% attack, nothing-at-stake problem together with improvements for the scalability issues in current blockchains. R. Belchior et al. in \cite{2005.14282} study cryptocurrency-directed interoperability approaches, blockchain Engines and blockchain connectors, providing a holistic overview of blockchain interoperability.

The integration of blockchain and 5G has become a mainstream trend, D. C. Nguyen et al. in \cite{Blockchain5Gandbeyondnetworks} provide  the latest survey on the integration of blockchain with 5G networks and other networks. It gives an extensive discussion about the potential of blockchain for enabling  key technologies of 5G, and further explores and analyzes the opportunities that blockchain may give important 5G services.  G. Yu et al. in \cite{8954616} study the sharding problem in the blockchain, mainly including providing detailed comparison and quantitative evaluation of the main sharding mechanisms, as well as analysis of the characteristics and limitations of existing solutions. Moreover, by enabling the integration of blockchain and other advanced technologies, some of work explore potential applications and research challenges in IoT \cite{Ferrag2019,platformsforIoTuse-cases}, smart city \cite{Xie2019}, cloud computing \cite{Gai2020}, edge computing \cite{Yang2019}, and fog computing\cite{9104970}.

From the perspective of mathematical tools,  Z. Liu et al. in \cite{Liu2019} provide reviews and analyses using game theory in detail to deal with a variety of problems regarding security, mining management and blockchain applications. However, some classic mathematical methods used in blockchain are not limited to game theory, but also include machine learning, cryptography, and so on. Moreover, a survey which provides a comprehensive overview on theoretical model, network service and management, and application for blockchain system on a methodology perspective is even more needed.

\subsection{Motivation}
The existing surveys  mainly focus on blockchain architecture, blockchain protocol algorithm, the integration of blockchain with other network technologies, etc. Besides, to the best of our knowledge, there is very few survey provides comprehensive reviews to elaborate the impact of communication networks and computing from methodology perspective. However, as a system science, blockchain requires researchers with multidisciplinary knowledge background in communication, network, computing and cryptography disciplines to understand the operating environment and   principles of blockchain from different dimensions, so as to make rational use of blockchain technology to improve innovative service applications in 5G/B5G, IoT and other fields.

This motivates us to review the state of the art of blockchain, to reveal the reason why a blockchain system needs communication, network and computing and how them interact to each other. Furthermore, to facilitate the deployment of blockchain-based applications, how to carry out mathematical methods to define a specific blockchain problem is a concern that many researchers deserve to pay attention to, but there is no survey specifically discussing the use of blockchain from the perspective of methodology such as stochastic process, cryptography and so on.  Motivated by this limitation, different from the previous surveys that focus on blockchain application such as IoT, smart city and edge computing or blockchain technology such as security and privacy, consensus algorithm and resource management, we overview the theoretical model for blockchain fundamentals understanding, and the design of network services for blockchain-based mechanisms and algorithms from a methodology perspective involving of stochastic process, game theory, optimization, machine learning, and cryptography, as well as summarizing the advantages and limitations of these methods.
%we introduce some classical mathematical methodologies in the blockchain running process and the application of  blockchain protocol/algorithm design, including stochastic process, game theory, optimization, machine learning, and cryptography, and summarize the advantages and limitations of these methods.
Moreover, this paper reviews some case studies and challenges applying blockchain in   IoT, providing feasible guidance and potential research problems.

%motivation部分要重写,重点考虑1:现有survey的局限性,区块链是一个系统科学,需要研究人员具备通信、网络、计算等多学科的知识背景,从不同维度理解区块链的运行环境和运作原理,才能在5G/B5G,iot 等多个领域合理利用区块链技术提高服务、创新应用；2现有survey主要都是以topic：区块链架构、区块链协议算法、区块链MEC 等等为导向,我们以方法论角度来探讨,帮助研究人员在各自熟悉的研究领域选择合适的研究方法,并且讨论了各种方法的实用性、优势和缺陷,（类似的survey 我记得就只有一篇bc+game的,你要重点review 和对比）

The organization of this article is as follows. Section II discusses the research activity, challenges and roadmap of  blockchain system. Section III-VII analysis the application of stochastic processes, game theory, optimization theory, machine learning, and cryptography for mining pool management, security strategy, resource management and so on in blockchain, respectively. Section VIII discusses main issues of blockchain and its improvements. Finally, Section IX concludes this paper.

\section{Research Activity, Challenge and Roadmap}
%Recently, lots of work have been done on  the application of blockchain and the improvement of network system performance, which involves blockchain-based function design such as consensus protocol, incentive mechanism and smart contract in blockchain layer and commutation, network and computing scheduling, management and optimization in infrastructure layer.

Recently, lots of work have been done on the application of blockchain and the improvement of network system performance, involving of the blockchain-based functions design and network optimization. The blockchain-based functions such as consensus protocol, incentive mechanism and smart contract have significant impacts on the reliability, efficiency and scalability of the blockchain system \cite{WireslessResouce}. However, the design of blockchain-based functions also faces challenges due to storage constraints, computing overhead and delay constraints. Moreover, blockchain is seen as a potential  technology to improve network system performance,  furthermore,  blockchain and the   network system especially for wireless network system interact and support each other \cite{2003.13083}. Specifically, on the one hand, with the advent of the 5G era, we are about to enter a highly dynamic wireless connected digital society mainly composed of massive wireless devices. Sequentially,  the   explosive information will likely be exchanged through wireless networks. However, the scarcity of wireless spectrum resources poses challenges for public in the safety and efficiency of resource/data management and sharing.
On the other hand, blockchain relies on frequent communication among consensus nodes to reach consensus. While the highly dynamic wireless network environment will  bring performance and security degradation to the communication within blockchain consensus nodes.

Therefore, the above-mentioned challenges   greatly hinder the safe and efficient application of blockchain in practical systems and weaken the contribution of blockchain to better improve the practical systems performances.
To cope with those challenges, it is first necessary to  understand the fundamentals  of blockchain,   the operating process of practical communication systems, as well as the impact of the   resources (e.g., communication, network computing resources, etc.) and other uncertain factors on the performance of blockchain. Thus, an accurate and efficient  theoretical model need to be established to analyze blockchain system performances and its influencing factors in essence.
Then the design of network services for blockchain-based  mechanisms and algorithms can be implemented, but which  often require the help of classical methodologies.
Finally, in order to facilitate the deployment of blockchain-based applications in IoT, Internet of Vehicles (IoV) and other   scenarios, it is also necessary to explore how to use mathematical methods to define a specific blockchain problem, and examine various constraints and application requirements in the practical systems from the perspective of methodology.
Therefore, methodology is the basis for studying the fundamentals, performances and applications of blockchain, and has been recognized. However, most of the existing review research directions focus on the combination of blockchain and other advanced technologies, the integration of blockchain and novel networks, and the security and privacy issues of blockchain technology, etc., with the advantage of providing  theoretical support for blockchain applications. But, it may also have certain limitations, which are affected by complex and uncertain factors in the practical system, and thus fail to further support the optimization of blockchain function design and network performance improvement. Therefore, it is necessary to systematically and comprehensively study the blockchain systems, technologies and applications from the perspective of methodology, and objectively evaluate their advantages and limitations.

In this Section, we first discuss the research activity classified into theoretical modeling for blockchain performance analysis, blockchain-based function design for network services, and blockchain-based solution for vertical applications. Next, we discuss the remaining challenge of blockchain system, and present the technical roadmap which illustrates the relationship in methodology, uses case and topic.

%此处是添加的介绍主要区块链项目的表和共识机制的性能比较图,还需要添加文字说明,由张龙师兄负责添加

\begin{figure}
  %\captionsetup{font={footnotesize}}
  \centering
  \includegraphics[width=9cm]{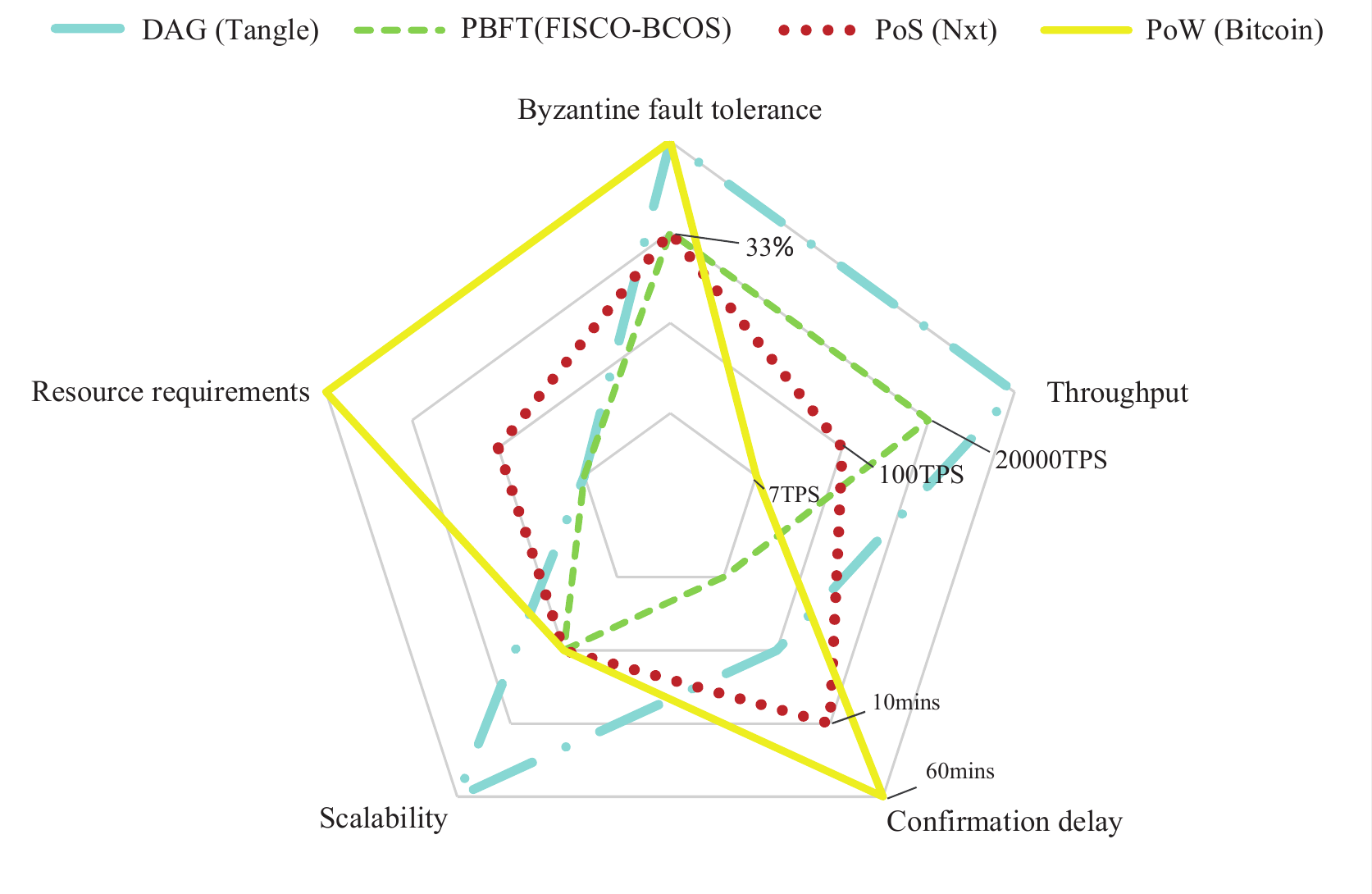}
  \caption{Multi-dimensional graph for performance comparison of typical consensus mechanisms in  blockchain}
  \label{fig:Multi-dimensionalgraph}
\end{figure}

%这部分内容我还是分为三大类,但名字更广泛,有利于深入浅出的讲解

\subsection{Research Activity}
\begin{table*}[]
  \caption{Major blockchain platforms}
  \label{Tab1}
  \centering
  \begin{tabular}{|c|c|c|c|c|c|c|c|}
    \hline
    Platforms                      & Bitcoin \cite{bitcoin}                       & Nxt \cite{nxt}                            & ETH \cite{ETH}                           & Hashgraph \cite{hashgraph}                    & IOTA \cite{IOTA}                           & \begin{tabular}[c]{@{}c@{}}Hyperledger \\ Fabric \cite{fabric} \end{tabular}  & EOS \cite{eos}                            \\ \hline
    \begin{tabular}[c]{@{}c@{}}Application \\ layer\end{tabular}  & \begin{tabular}[c]{@{}c@{}}Bitcoin \\ transaction\end{tabular}  & \begin{tabular}[c]{@{}c@{}}DAPP/Nxt \\ transaction\end{tabular}  & \begin{tabular}[c]{@{}c@{}}DAPP/ETH \\ transaction\end{tabular}  & \begin{tabular}[c]{@{}c@{}}Public \\ network\end{tabular}  & \begin{tabular}[c]{@{}c@{}}IOTA \\ transaction\end{tabular}  & \begin{tabular}[c]{@{}c@{}}Enterprise \\ block \\ application\end{tabular} & \begin{tabular}[c]{@{}c@{}}Operating \\ system\end{tabular} \\ \hline
    \begin{tabular}[c]{@{}c@{}}Programming \\ language\end{tabular} & JavaScript                     & Java                           & \begin{tabular}[c]{@{}c@{}}Solidity/\\ Serpent\end{tabular} & -                              & \begin{tabular}[c]{@{}c@{}}JavaScript/\\ Java/C\#/Go\end{tabular} & Go/Java                        & C++                            \\ \hline
    Data model                     & \begin{tabular}[c]{@{}c@{}}Transaction \\ model\end{tabular} & Account model                  & Account model                  & -                              & Account model                  & Account model                  & \begin{tabular}[c]{@{}c@{}}Account \\ model\end{tabular} \\ \hline
    \begin{tabular}[c]{@{}c@{}}Block \\ storage\end{tabular} & LevelDB                        & -                              & LevelDB                        & -                              & -                              & FileSystem                     & -                              \\ \hline
    \begin{tabular}[c]{@{}c@{}}Communication \\ protocol\end{tabular} & P2P                            & P2P                            & P2P                            & -                              & HTTP                           & P2P                            & P2P                            \\ \hline
    Category                       & \begin{tabular}[c]{@{}c@{}}Public \\ blockchain\end{tabular} & \begin{tabular}[c]{@{}c@{}}Private \\ blockchain\end{tabular} & \begin{tabular}[c]{@{}c@{}}Public \\ blockchain\end{tabular} & \begin{tabular}[c]{@{}c@{}}Consortium \\ blockchain/\\ Private \\ blockchain\end{tabular} & \begin{tabular}[c]{@{}c@{}}Public \\ blockchain\end{tabular} & \begin{tabular}[c]{@{}c@{}}Consortium \\ blockchain\end{tabular} & \begin{tabular}[c]{@{}c@{}}Consortium \\ blockchain\end{tabular} \\ \hline
    \begin{tabular}[c]{@{}c@{}}Consensus \\ algorithm\end{tabular} & PoW                            & PoS                            & PoW                            & BFT                            & Tangle                         & BFT                            & DPoS                           \\ \hline
    Architecture                   & \begin{tabular}[c]{@{}c@{}}Continuous \\ single chain \\ architecture\end{tabular} & \begin{tabular}[c]{@{}c@{}}Continuous \\ single chain\\ architecture\end{tabular} & \begin{tabular}[c]{@{}c@{}}Continuous \\ single chain \\ architecture\end{tabular} & \begin{tabular}[c]{@{}c@{}}Based on \\ DAG \\ architecture\end{tabular} & \begin{tabular}[c]{@{}c@{}}Based on \\ DAG \\ architecture\end{tabular} & \begin{tabular}[c]{@{}c@{}}Continuous \\ chain \\ architecture\end{tabular} & \begin{tabular}[c]{@{}c@{}}Continuous \\ chain \\ architecture\end{tabular} \\ \hline
    Solution                       & \begin{tabular}[c]{@{}c@{}}Computational \\ power \\ competition\end{tabular} & Coin age                       & \begin{tabular}[c]{@{}c@{}}Computational\\ power \\ competition\end{tabular} & \begin{tabular}[c]{@{}c@{}}Virtual \\ voting, \\ gossip \\ protocols\end{tabular} & \begin{tabular}[c]{@{}c@{}}Self-weight \\ and \\ cumulative \\ weight\end{tabular} & \begin{tabular}[c]{@{}c@{}}Gossip \\ protocols, \\ endorsement \\ and ordering\end{tabular} & \begin{tabular}[c]{@{}c@{}}Virtual \\ voting\end{tabular} \\ \hline
    Characteristic                 & \begin{tabular}[c]{@{}c@{}}Low \\ transaction \\ fee, high \\ security, low\\ network \\ resource\\ consumption\end{tabular} & \begin{tabular}[c]{@{}c@{}}Proof-of-\\ Stake\\ consensus, \\ universal \\ blockchain \\ framework, \\ decentralized\\ asset \\ exchange,\\ proven \\ stability\end{tabular} & \begin{tabular}[c]{@{}c@{}}Low \\ transaction\\ fee, high \\ security, low\\ network \\ resource\\ consumption\end{tabular} & \begin{tabular}[c]{@{}c@{}}Low \\ consensus \\ cost, high\\ security,\\ high \\ transaction \\ throughput, \\ low\\ confirmation\\ latency\end{tabular} & \begin{tabular}[c]{@{}c@{}}High \\ scalability, \\ low resource \\ requirements, \\ zero-fee \\ transactions, \\ secure data \\ transfer, \\ offline \\ transactions, \\ quantum immune\end{tabular} & \begin{tabular}[c]{@{}c@{}}High \\ scalability, \\ permission \\ control and \\ modular \\ architecture\end{tabular} & \begin{tabular}[c]{@{}c@{}}Flexible, \\ scalable, \\ user-\\ friendly\end{tabular} \\ \hline
    \begin{tabular}[c]{@{}c@{}}Open source \\ address\end{tabular} & \begin{tabular}[c]{@{}c@{}}https://bit\\ coincore.org/\\ en/download/\end{tabular} & \begin{tabular}[c]{@{}c@{}}https://bitbuc\\ ket.org/Jeluri\\ da/nxt/src/\\ master/\end{tabular} & \begin{tabular}[c]{@{}c@{}}https://geth.\\ ethereum.org/\\ downloads/\end{tabular} & -                              & \begin{tabular}[c]{@{}c@{}}https://github.\\ com/iotaledger/\\ iota.js\end{tabular} & \begin{tabular}[c]{@{}c@{}}https://github.\\ com/hyperledger\\ /fabric\end{tabular} & \begin{tabular}[c]{@{}c@{}}https://\\ github.com\\ /EOSI\\ O/eos\end{tabular} \\ \hline
  \end{tabular}
\end{table*}

\subsubsection{Theoretical modeling for blockchain performance analysis}

%经典的区块链一致性的理论基础来源于拜占庭将军问题,然而,具体的区块链系统和相应的协议算法如何确保拜占庭容错,以及对应的通信、网络和计算开销,算法复杂度分析、收敛速度等等一系列重要问题缺乏系统性的分析。模拟区块链系统的运作过程,从而帮助研究人员理解区块链核心协议算法的运作原则,关键步骤,从而建立精确有效的理论模型分析区块链系统性能和安全性,提供可靠的理论指导,获取一致性达成的必要条件、达成时延和开销、区块链处理能力等系统性能,以及对恶意攻击的容忍度等重要理论指标。以此为理论基础,理解区块链运行环境、协议算法流程、用户行为、恶意攻击手段及其影响,从而为区块链系统的优化设计,基于区块链的网络服务设计,基于区块链在xxx

%To investigate the performance of a blockchain system, existing projects and researches mainly focus on the theoretically analyzing the impact of stochastic variables on throughput, delay, cost, etc., designing various consensus protocols for specific application, and decoupling the traditional centralized network architecture for security and scalability. Nevertheless, the research on blockchain system considering resource-constrained and uncertain communication, network and computation is still at its infancy stage. To this end, a theoretical model that can capture the characteristics of complex and dynamic scenarios should be further designed to analyze the performance of blockchain system in resource-constrained and uncertain communication, network and computation.
The establishment of accurate and effective theoretical model is very important to study the system performance such as the necessary conditions for achieving consistency, delay and cost of achieving consistency, and processing capacity of blockchain, which can provide reliable theoretical guidance for analyzing the performance of blockchain system. Existing projects and studies on system performance focus on he theoretically analyzing in stochastic variable analysis, design of consensus protocols for specific applications, and decoupling of the traditional centralized network architectures for security and scalability. Specifically, in the practical blockchain system, there are random behaviors such as block generation time, confirmation delay and chain growth rate. Thus, the random variable analysis is often used to observe its impact on the throughput, delay and cost of the blockchain system, so as to help researchers effectively and accurately study the running process of the blockchain \cite{1article, 8818339,2002.02567}.
%In addition， Consensus plays a key role in the blockchain, which largely determines blockchain system security bound and performance.
In addition, the innovation of different types of consistent algorithms which have different characteristics and serve different purposes has great significance to the security and efficiency of blockchain systems, and thus attracted the attention of many scholars \cite{ConsensusAlgorithms}. Furthermore, in order to cope with the shortcomings of single point of failure caused by traditional centralized network, the blockchain technology is used to reshape the  network architecture to improve the network scalability, reliability and positivity, and the interaction between blockchain and network can be observed through the fusion of blockchain and  networks \cite{8598796}.

However, in  practical blockchain systems, the factors affecting its performance are rich and complex. Therefore, in order to better perform blockchain performance analysis, in addition to the aspects mentioned above, resource constraints and uncertain aspects of communication, network and computing need to be considered. Thus, it is also necessary to further design a theoretical model that can capture the characteristics of complex dynamic scenarios for the optimal design of the blockchain system, and the network service design based on the blockchain, to provide the necessary theoretical guidance and design ideas for technological innovation and breakthroughs.

%第二段,介绍现在主要的modelling工作

\subsubsection{Blockchain-based function for network services}

\begin{figure*}
	\centering
	\includegraphics[width=0.7\linewidth]{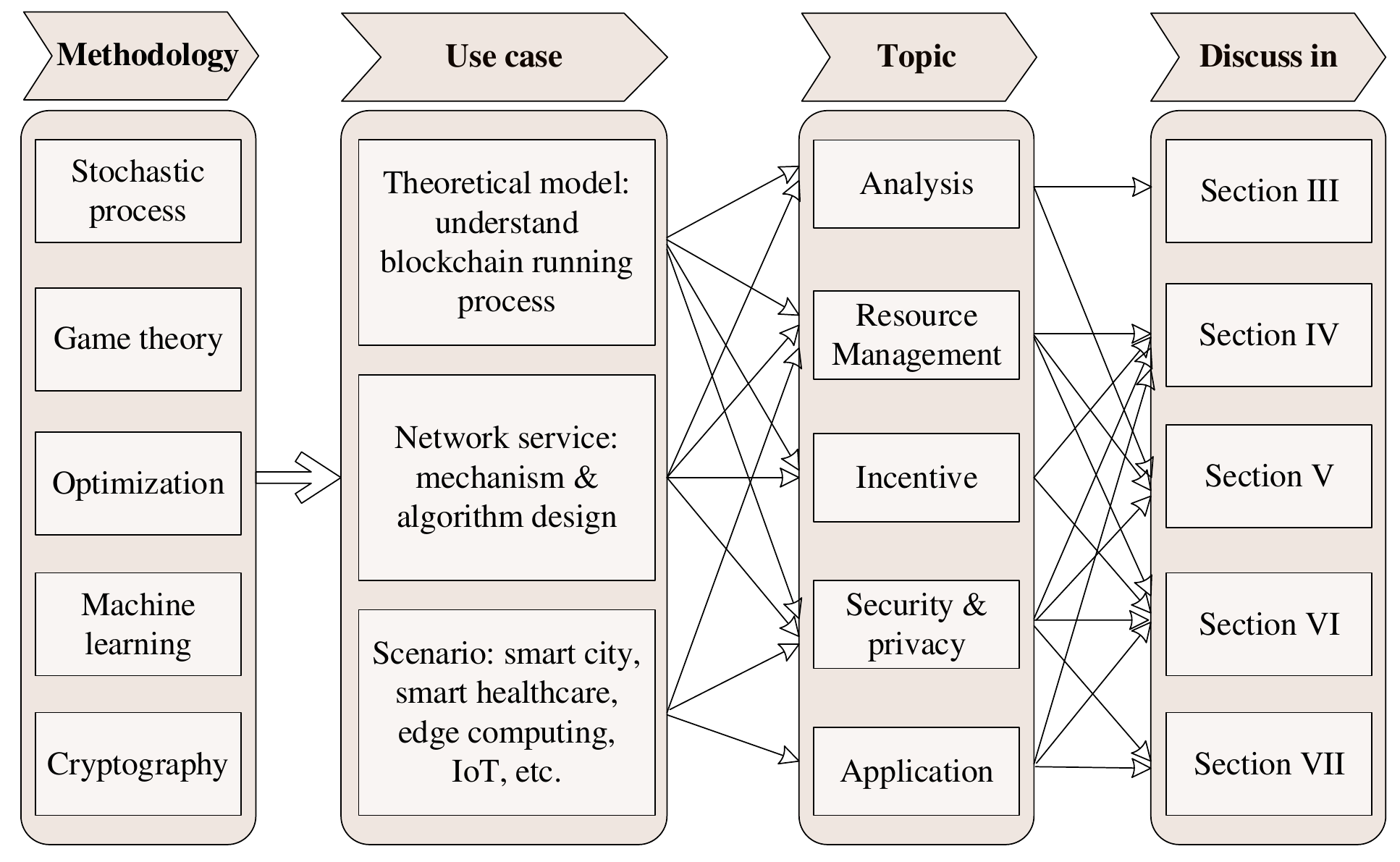}
	\caption[ ]{The technical roadmap on the methodology perspective}
	\label{fig:bc-survey-roadmap}
\end{figure*}

%\begin{figure*}[t]
  %\captionsetup{font={footnotesize}}	
 % \centering
%  \includegraphics[width=13cm]{picture/chapter2/BC-survey-roadmap}
%  \caption{The technical roadmap on the methodology perspective}
%  \label{fig:bc-survey-roadmap}
%\end{figure*}
Relying on mechanism and algorithm design, blockchain can achieve various functions and provide network services in a distributed way, including management, incentive, security \& privacy, and resource allocation. In addition to the initial financial service, more research related to blockchain services is being concentrated on specific areas relevanted  to network services, such as  public and social services \cite{networkservice1}, cloud services \cite{7990130}, and other Internet services.
For distributed systems with blockchain participation, the consistency of state among nodes is a key to ensure the integrity and security.  Importantly,  blockchain system exactly uses the consensus algorithm  to achieve the consistency of state among nodes, which plays a key and irreplaceable role.  This fact has fuelled the consensus algorithm is the core component that directly dictates how such a system behaves and the performance it can achieve.
So that, the consensus mechanism is the basis for blockchain to establish trust and agreements without the participation of third parties. Generally speaking, consensus mechanism is a system that can constrain each decentralized node in the decentralized network, maintain the operation order and fairness of the system, and enable each unrelated node to verify and confirm the data in the network, so as to generate trust and reach consensus. consequently, consensus plays a key role in the blockchain, which largely determines blockchain system security bound and performance.

To establish consistency, various consensus algorithms for different projects are proposed. Most consensus algorithms are originated from typical Proof-of-Work (PoW), Proof of Stake (PoS), Practical Byzantine Fault Tolerance (PBFT) and Raft, and the above algorithms have their own advantages and drawbacks in throughput, confirmation latency, security, transaction fee and tolerate fault. For example, PoW and PoS are completely decentralized consensus algorithms, but they require much meaningful computation to win the competition \cite{8428402}. PBFT and Raft \cite{2003.13083}\cite{8982036} have good performance, but they are suitable for private networks. Besides, the performance of Raft depends on honest nodes and cannot solve BFT. For some typical consensus algorithms, a multi-dimensional performance comparison is shown in Fig. \ref{fig:Multi-dimensionalgraph}.

%第二段,介绍基于共识达成的network service设计
\subsubsection{Blockchain-based solution for vertical applications}

Owning to the benefits of building trust, reducing cost and accelerating transactions,
blockchain  technology is expanding to other areas including IoV, the industry 4.0, smart homes, artificial intelligence integrated services, and such \cite{8548560,chapter2research1,articlesmatrhome,2003.13083}.
%作为行业应用的重要承载，区块链平台平台建设越来越受到各大区块链企业的重视，现有主流的去开了平台分析如表1所示。,,,,.这些限制限制了区块链平台发展与行业应用落地。
In addition, as an important carrier of industrial applications, the construction of blockchain platform has attracted more and more attention from major blockchain enterprises. The analysis of existing mainstream platforms is shown in Table \ref{Tab1}.
\iffalse blockchain technology has a huge potential to promote the development in IoT \cite{chapter2research1}, edge computing \cite{chapter2research3}, and smart city \cite{chapter2research2}, the industry 4.0, smart homes, artificial intelligence integrated services.\fi
However, as we discussed before, the performance of the mainstream blockchain technologies is limited by  communication, network and computation, resulting in some shortcoming such as high confirmation delay, low throughput, intensive computation, and redundant storage.
%Moreover, These limitations restrict blockchain-enabled applications in constrained and uncertain environments.
Moreover, these limitations restrict the development of blockchain platforms and the landing of industrial applications.
Naturally, it is necessary to propose feasible architecture and workflow to support blockchain platform construction and application development with satisfactory communication, network and computation capabilities.
%Nowadays, some studies have proposed to integrate blockchain with edge cloud and fog computing to deal with complex computing task and improve system performance.

%第二段,区块链解决应用问题的研究

\subsection{Challenges}

\subsubsection{Theoretical model}

Targeted at the problem of scalability, transaction throughput, latency and security, etc, various algorithms and protocols are designed in existing researches on blockchain \cite{chapter2challenge4} \cite{chapter2challenge5}. %Aiming to break through Bitcoin's inherent scalability limits, \cite{chapter2challenge4} has presented Block-NG, a new blockchain protocol. To support secure crosschain transactions, the authors in \cite{chapter2challenge5} have proposed a series of techniques to ensure synchronization and atomicity of crosschain transactions.
In these researches, performance analysis are conducted  through experimental emulations or application implements. However, without a complete theoretical model,  it is difficult to accurately find out the impact of user behaviors and system parameters on the performance and security, thus leading to the absence of the precise theoretical indicators, such as performance boundaries, tolerance to malicious attacks and so on. Importantly, the lack of theoretical foundation cannot provide theoretical guidance for the rational application or technological breakthrough of the blockchain, which will further limit the development of the blockchain.
Therefore, facing the rapid popularization of blockchain application, to make up for the relative backwardness of the theoretical basis, an accurate and easily extensible set of mathematical theoretical models is necessary to be established.

%（2）技术融合,解决现有区块链与智能化无线网络之间架构、算法和协议不能完全匹配的矛盾。将区块链技术应用于智能化无线网络环境,需要从两个方面考虑：首先,区块链和智能化无线网络的设计初衷不尽相同,前者的本质是一个去中心化的数据库,后者的核心是一个实现万物互联的信息承载体,因此,需要设计一套全新的体系架构和流程,将两种技术有效结合,以智能化无线网络实现数据的感知、控制、传输、计算和汇聚,以区块链对数据进行共识、处理、存储和保护。其次,现有的区块链算法并不是针对智能化无线网络所设计,在链路接入、资源调度等方面均不能完全支持不同的区块链算法对通信、网络和计算资源的特定要求。同时,主流的区块链技术往往受到算力、时延、复杂度、交易费用、通信开销和网络流量的影响,难以直接应用于智能化无线网络。所以,这就需要将现有区块链和智能化无线网络的算法协议进行评估,重新设计,使得两者在保持各自技术优势的前提下,实现不同算法和协议之间的匹配,推进两种技术的有效融合。

%\subsubsection{Network service and management}
\subsubsection{Technique fusion}
Known as a secure, reliable and transparent distributed ledger, blockchain has been heralded as a technology that can energize many scenarios for practical systems, such as spectrum management, Industrial Internet of Things (IIoT) and so on %[Edge Intelligence and Blockchain Empowered 5G Beyond for the Industrial Internet of Things]
\cite{chapter2challenge1} %[On the Application of Blockchains to Spectrum Management]
\cite{chapter2challenge2}. However, how to integrate blockchain into practical systems especially into wireless networks, while ensuring its security and performance, is still challenging. Designed for different purposes, the essence of blockchain is data storage and consensus while that of wireless networks is for data perception, transmission and aggregation. It is necessary to propose a new set of architectures and processes that can effectively combine them. Furthermore, as the mainstream blockchain protocols, PoW is of high resource consumption, direct acyclic graph's performance is highly influenced by information load and byzantine fault tolerance is of high communication complexity. These characteristics form the obstacle for them to be applied to practical system, which is with complex communication environment, limited resources and diversified service requirements. Therefore, faced with the mismatches from communication, network and computation, a dedicated blockchain protocol for practical system is necessary to be designed.

%（3）联合优化,解决现有区块链性能和智能化无线网络应用环境和业务需求之间的矛盾。现有的主流区块链技术以证明方法（如工作量证明）,投票方法（如哈希图）和容错方法（如拜占庭容错）为代表,这些区块链技术虽各具特点和优势,但都不能完全满足智能化无线网络的实际需求。证明方法往往资源消耗大、时延高、吞吐量低和交易费用昂贵等问题,不适用于资源受限且业务量多样化的智能化无线网络应用场景；基于有向非环图的哈希图方法虽然可以解决上述问题,但其性能极易受当前网络负载大小的影响,智能化无线网络业务流的实时变化会导致强烈且不可控的性能波动；拜占庭容错方法会产生大量的广播消息,为智能化无线网络带来不可忽视的系统负荷和额外开销。因此,急需面向大规模部署、资源受限、网络异构、业务多样、移动时变、链接不确定性环境下的专用区块链技术,满足快速共识、并发处理、高吞吐量、低时延、低费用、易于扩展和健壮安全等应用需求。

%另外,随着智能化无线网络的快速发展和普及,不同用户在资源、数据、成本等方面具有各自的优势,需要协调不同用户的自身收益,使其到达个体最优的同时,促使用户帮助区块链系统实现一致性。因此,考虑这种复杂动态的竞争博弈关系,需要建立一种分布式的高效协同与激励机制,使系统快速达到均衡状态,用户个体利益和系统性能之间维持动态平衡。

%同时,在系统运行过程中,还必须维护用户对自身数据的所有权并进行高效的安全和隐私保护,防止、识别不同类型的攻击。

%再者,考虑智能化无线网络的资源有限性可能导致区块链的性能瓶颈,可以采取雾计算等新一代移动网络技术,帮助区块链处理复杂计算任务,提高系统性能。

%为此,就需要将区块链物联网与其他先进技术相互匹配与联合优化,从干扰协调、链路接入、资源调度、无缝连接等多维角度出发,对智能化无线网络用户、区块链功能节点、蜂窝基站记性联合部署,实现区块链和网络之间的性能平衡,消除单方面的瓶颈容量,最终提高系统总体性能。

\subsubsection{Joint optimization}

Performance and security of wireless network based blockchain are not only affected by the designed blockchain protocol, but also by application environment and business requirements of the intelligent wireless network.
%the coexisting Radio Access Network (RAN) technologies.
For example, as to Blockchain-Enabled MEC Systems %[Joint Optimization of Radio and Computational Resources Allocation in Blockchain-Enabled Mobile Edge Computing Systems]
\cite{chapter2challenge3}, not only delay/time to finality (DTF) for the blockchain system but also energy consumption for the MEC system, are the performance metrics that need to be considered. Then, to avoid the sub-optimal performance resulted in by separate optimizations, a joint optimization problem can be formulated to achieve the optimal trade-off between these two performance metrics. Generally, when it comes to the matching of blockchain and other technologies, from the perspective of multi-dimensional performance requirements, in order to eliminate unilateral bottlenecks and ultimately achieve the overall system performance, the joint optimization is necessary to be considered.

\subsection{Roadmap on the Methodology Perspective}
%Different from the previous surveys that focus on blockchain application such as IoT, smart city and edge computing \cite{Ferrag2019}\cite{platformsforIoTuse-cases}\cite{Xie2019}\cite{Gai2020}\cite{Yang2019} or blockchain technology such as security and privacy, consensus algorithm and resource management \cite{8369416}\cite{Salman2019}\cite{Saad2020}\cite{ConsensusAlgorithms}\cite{Wang2019}\cite{Liu2020}\cite{Xiao2020},
This article aims to provide a comprehensive overview on theoretical model, network service and management, and application for blockchain system on a methodology perspective. In order to illustrate the relationship in methodology, use cases  and topic, a technical roadmap is shown in Fig. \ref{fig:bc-survey-roadmap}. First, we introduce the calssic methodology to investigate running process of blockchain to formulate and analyze. Then illustrate the use cases for blockchain theoretical model, blockchain technology design enabling network service, and  how to use blockchain as the fundamental of industry vertical applications effectively. Accordingly, we review the state of the art of blockchain system for performance and security analysis, resource management, incentive mechanism, ect, and some remaining problems  are  summarized and discussed.

%First, we introduce the method to investigate running process of blockchain to formulate and analyze, then illustrate the approach for blockchain technology design enabling network service, and finally, show how to use blockchain as the fundamental of industry vertical applications effectively. Accordingly, we review the state-of-the-art of blockchain system for performance and security analysis, network management, incentive mechanism,

\section{Stochastic Process in Blockchain}

Stochastic Process \cite{chapter3Introduction1} is a mathematical method to establish system model and analyze the performance index in uncertain environment, it is a set of time-dependent random variables used to describe the system state at a specific time. In communications, especially in wireless scenario, the complex environment can be abstracted as stochastic process for analysis in a mathematical manner. As well known, Markov process has been widely used to describe the common random process. Generally, if the current system state is determined, the trend of system state in the future would be also known, no matter what the system state is in the past. Therefore, the Markov process is an important and effective mathematical tool to predict the system state, user action and network performance. Meanwhile, stochastic geometry \cite{chapter3Introduction2} is another mathematical tool widely used in the modeling and analysis of communications and networks \cite{chapter3Introduction3}. In communication networks, the actual network nodes and spatial locations can be formulated as random point processes, such as Homogeneous Poisson Point Process (HPPP), which is to eliminate the randomness by traversal to analyze the system performance and provide design ideas theoretically.

\subsection{Model Briefs}

\par In practical blockchain system, there exist many random behaviors such as block generation time, confirmation delay, chain growth rate, and forking. As mentioned before, it is a nature design to adopt stochastic process to model these random behaviors systematically, which can help us to formulate the blockchain operation process effectively and accurately. Next, we discuss how to use stochastic process in blockchain system on the typical issues of consensus, security and deployment.
In order to formulate the consensus process in blockchain system as a stochastic process, it is necessary to define a metric (like the cumulative block in PoW, or the cumulative weight in DAG) to indicate the consensus state at the different time. Moreover, if it satisfies Markov properties, the consensus process can be formulated as a Markov chain model with one-step transition probability
$P=P\{X_{n+1}=i_{n+1}|X_{n+1}=i_{n}\}$.
%%%
%此处插入第一图
\begin{figure}[H]
  %\centering
  \includegraphics[width=3.5in]{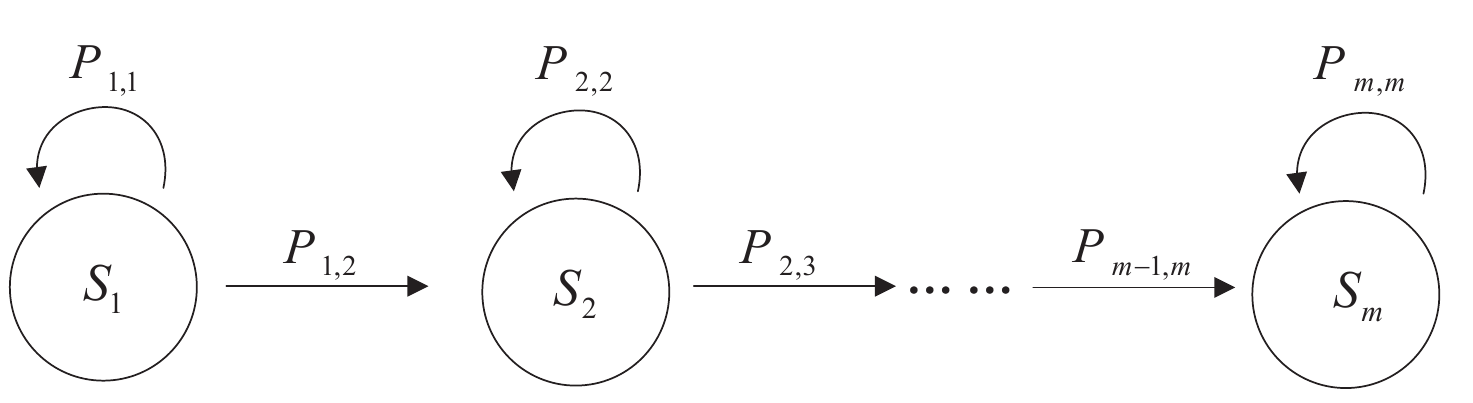}
  % where an .eps filename suffix will be assumed under latex,
  % and a .pdf suffix will be assumed for pdflatex; or what has been declared
  % via \DeclareGraphicsExtensions.
  \caption{The Stochastic Process in consensus process of the Blockchain. }
  \label{fig-3-1}
\end{figure}
\par As shown in Fig. \ref{fig-3-1}, we use Markov chain to model the consensus process. In this process, $S_i (i=1,2...,m)$ represents the state in the consensus process, and $P_{i,j}$ is the one-step transition probability in the Markov chain mode. Accordingly, we can learn the consensus process and gradually understand the inner-action. Furthermore, it is also useful to analyze the malicious forking attack for security.

\par
\begin{figure}[ht]
  %\centering
  \includegraphics[width=3.5in]{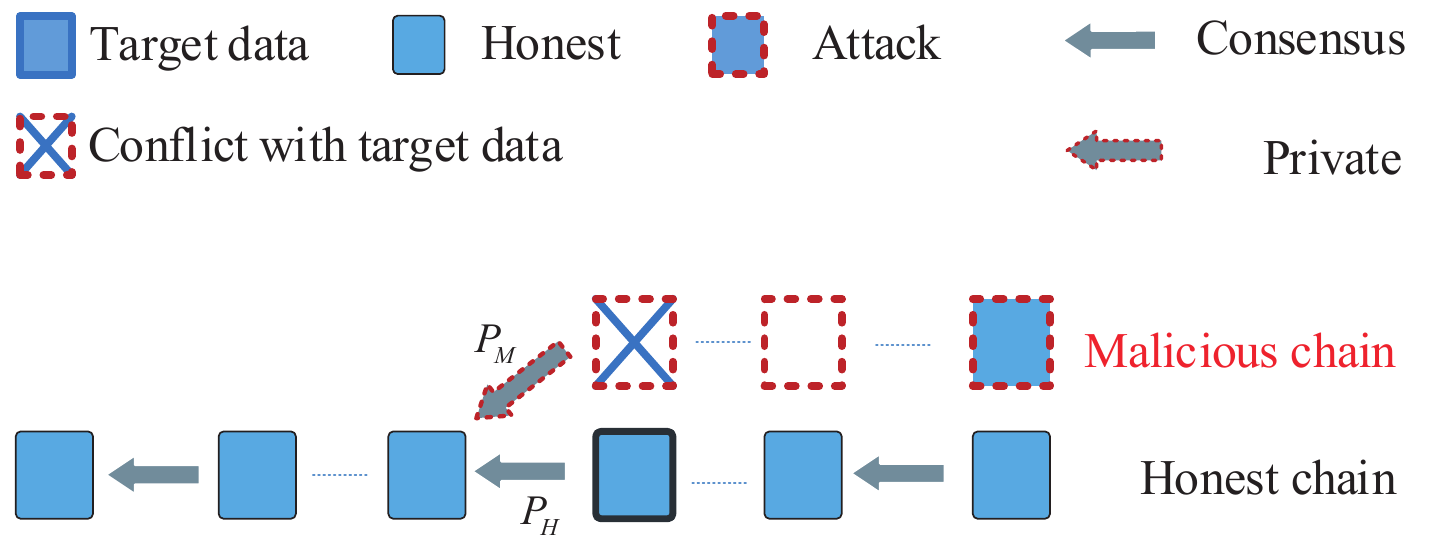}
  % where an .eps filename suffix will be assumed under latex,
  % and a .pdf suffix will be assumed for pdflatex; or what has been declared
  % via \DeclareGraphicsExtensions.
  \caption{The Stochastic Process in the block generation of the Blockchain.  }
  \label{fig-3-2}
\end{figure}

 As the most famous consensus process, the PoW-based mining task proposed by bitcoin is a competition among minors, where the winner has the right to generate a new block to obtain an amount of reward. For malicious purposes like double-spending in the blockchain, forking attack launched by the malicious node which generates blocks to build a parasite chain privately, it would succeed if the parasite chain is longer than the main chain built by the honest node due to Longest-Chain-Rule (LCR). Indeed, this attack can be treated as a competition between honest and malicious nodes. Particularly, without any malicious node, this competition is also the consensus process formulated previously. Therefore, the competition for block generation can be modeled as a Poisson process to study the relationship between honest and malicious nodes, which is shown in Fig. \ref{fig-3-2}. The malicious nodes compete against the honest nodes to generate the block. The node which has the biggest probability  will have the right to generate the block. Accordingly, the factors affecting the vulnerability of blockchain can be known, and the successful probability of malicious node can be determined. As a result, the theoretical insight can be provided to resist forking attack, optimize consensus mechanisms, and improve network security.

%%%
%此处插入第二图
\par Like the classic base station deployment problem in heterogeneous networks, deployment of blockchain funtion node (or called as full node in some literatures)  can be solved using a stochastic method/stochastic methods in the same manner. The blockchain system is decentralized which is composed of multiple distributed blockchain nodes, their geographical distribution can be modeled as a Homogeneous Poisson Point Process (HPPP). In this way, we can construct an effective blockchain system architecture, and analyze the impact of blockchain node distribution on the communication throughput, SNR, security and other network metrics. Therefore, a valuable theoretical guidance is developed to optimize the blockchain node deployment  in order to imporve the system performance.

%%%
%此处插入第三图
%应冯老师要求,删除此图
\iffalse
\begin{comment}
\begin{figure}[H]
  %\centering
  \includegraphics[width=3in]{picture/chapter3/chapter3-3-eps-converted-to.pdf}
  % where an .eps filename suffix will be assumed under latex,
  % and a .pdf suffix will be assumed for pdflatex; or what has been declared
  % via \DeclareGraphicsExtensions.
  \caption{their geographical distribution of users in blockchain can be modeled as a Homogeneous Poisson Point Process (HPPP).  The blockchain users are associated with the neareast full nodes.}
  \label{fig-3-3}
\end{figure}
\end{comment}
\fi

\subsection{Case Study}

\par
\begin{figure}[h]
  %\centering
  \includegraphics[width=3.5in]{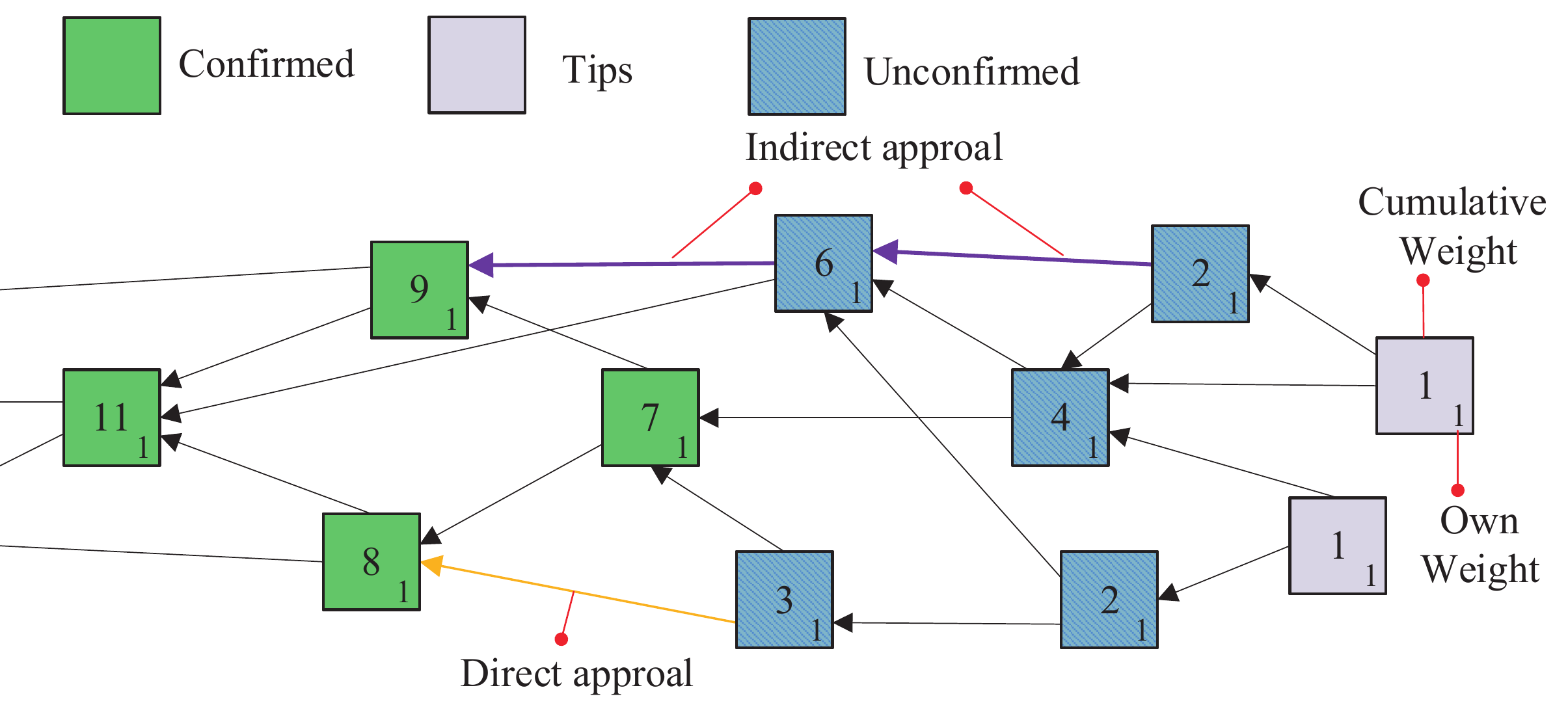}
  % where an .eps filename suffix will be assumed under latex,
  % and a .pdf suffix will be assumed for pdflatex; or what has been declared
  % via \DeclareGraphicsExtensions.
  \caption{The structure of DAG consensus \cite{chapter3CaseStudy}.}
  \label{fig-3-DAG}
\end{figure}

 In this section, we will use our previous work \cite{chapter3CaseStudy}
%[Direct Acyclic Graph based Blockchain for Internet of Things: Performance and Security Analysis] [Yixin Li]\cite{chapter3CaseStudy}
as an example to introduce how to model the classic blockchain problem as a stochastic process. As shown in Fig. \ref{fig-3-DAG}, Tangle is a DAG based distributed ledger for recording transactions.
\par DAG consensus is a typical voting mechanism to accumulate weight for consensus, which allows the new transaction to randomly select two unselected transactions that are called as tips. Therefore, according to the random selection in DAG consensus,  selected times of the observed transaction within the time period $[0,t]$ is a random variable of $t$ based on the assumption that the arrival time of the observed transaction is 0.
%Assuming that the arrival time of the observed transaction is 0, according to the random selection in DAG consensus (it is a typical voting mechanism to accumulate weight for consensus, which allows the new transaction to select two unselected transactions called as tips randomly $[x]$), selected times of the observed transaction within the time period $[0,t]$ is a random variable of $t$.
Thus, the cumulative weight of the observed transaction is its own weight plus the overall number of transactions which select it (the weight of each transaction is assumed to be 1). Therefore, the cumulative weight $ W (t) $ is a random variable with time $t$ ($ \{W (t), t \in [0, \infty]\}$), which is a stochastic process.  Meanwhile, $ L(t)$ is the number of tips at time $t$, and it is also a stochastic process because it is affected by the random selection as well as $W(t)$.
%%%
%此处解释什么是DAG

\par In summary,    the future states of $L(t+1)$ and $W(t+1)$ are determined by their current states ($L(t)$ and $W(t)$) only, which satisfies Markov properties. Therefore, this consensus process can be formulated as a Markov chain.  Due to the consideration that the new transaction arrives slowly resulting in a low network load, the system state is defined as $\{W (t), L (t)\}$ that is modeled as a discrete Markov chain $\{W (k), L ( k)\}, k = 0,1,2, ..., \infty$.
\par When a new transaction $x$ arrives, the   change in system state can be expressed as
\begin{align}
W\left( {k + 1} \right) = \left\{ {\begin{array}{*{20}{c}}
{W\left( k \right)} &{{a_x} = 1};\\
{W\left( k \right){\rm{ + 1}}} &{{a_x} = 0}.
\end{array}} \right.
\end{align}
\begin{align}
L(k+1)=L(k)-1.
\end{align}
where the $a_x=1$ stands for the situation that the observed transaction has been approved by an incoming new transaction,  and $a_x=0$ stands for the situation that the observed transaction has not been approved. Since the new transaction should select two unselected transactions randomly, thus (3.2) indicates that the new transaction replaces two unselected transactions as a new one.
%$W (k+1)=W(k)$ stands for the situation that the observed transaction has been approved by an incoming new transaction,  and $W (k+1)=W(k)+1$  stands for the situation that the observed transaction has not been approved. Since the new transaction should select two unselected transactions randomly, thus $L(k+1)=L(k)-1$ indicates that the new transaction instead of two unselected transactions as a new one.

\begin{figure}[H]
  %\centering
 \includegraphics[width=3.5in]{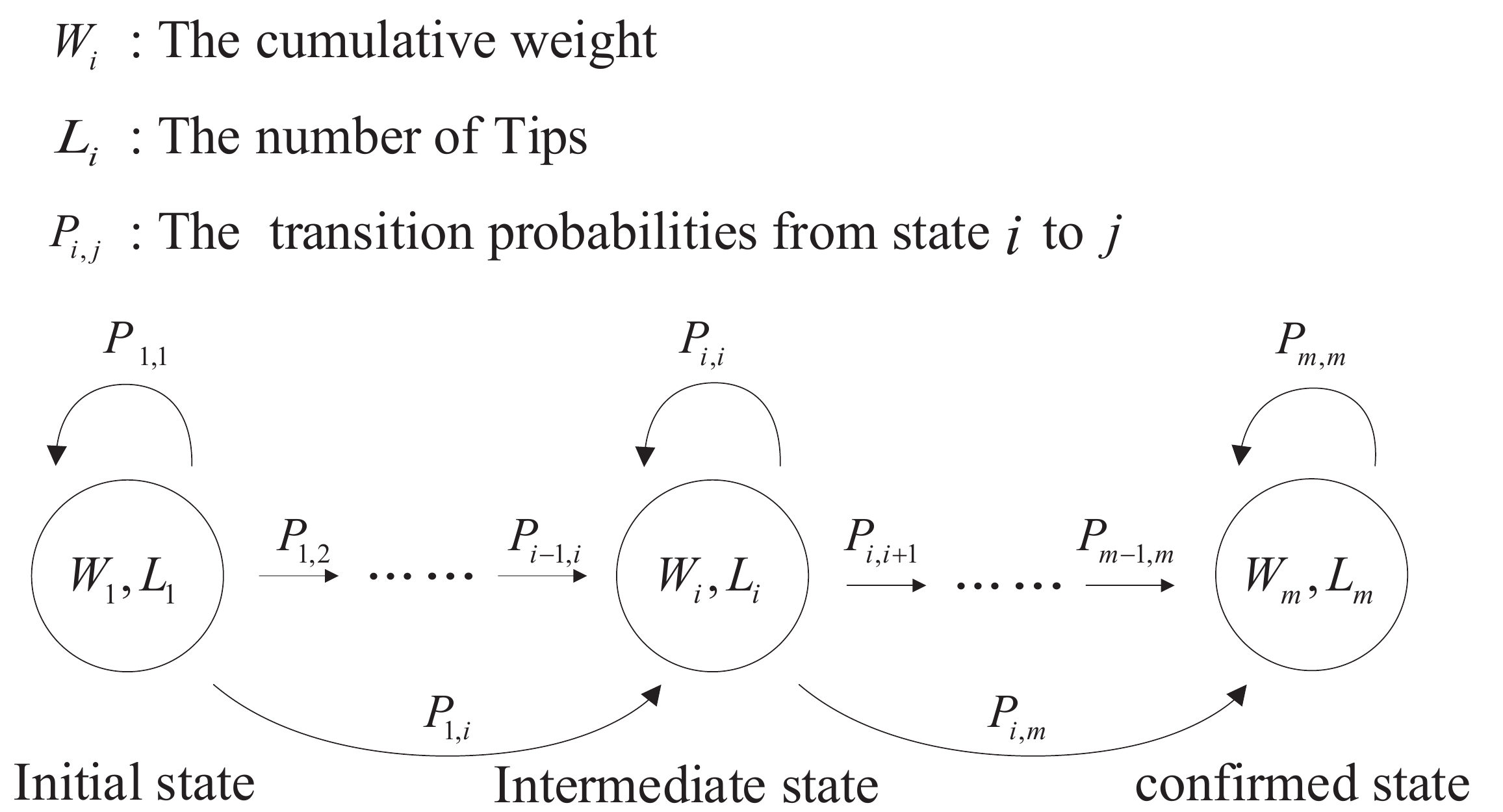} 
 % \includegraphics[width=3.5in]{picture/chapter3/3-CaseStudy}
  % where an .eps filename suffix will be assumed under latex,
  % and a .pdf suffix will be assumed for pdflatex; or what has been declared
  % via \DeclareGraphicsExtensions.
  \caption{The  Markov chain for DAG-baed consensus process.}
  \label{fig-3-CaseStudy}
\end{figure}

Therefore, the corresponding one-step transition probabilities and Markov chain can be shown as Fig. \ref{fig-3-CaseStudy}.
%%%
%此处插入case study 插图
\par For example,  if the network is from high to low with Unsteady State, the one-step transition probabilities can be expressed as:

\iffalse\begin{small}
  \begin{equation}
    \begin{cases}
       & P\{i+1,j-1|i,j\}=2/j,\quad i=1,2,...,L_n-1;j=2,3,\dots,L_h, \\%mathbb花体字符
       & P\{i,j-1|i,j\}=1-2/j,\quad i=1,2,...,L_n-1;j=2,3,\dots,L_h	, \\
       & P\{i+1,1|i,j\}=1,i=2,3,...,\infty;\quad j=1.\
    \end{cases} \notag
  \end{equation}
\end{small}\fi

\begin{footnotesize}
\begin{align}
\left\{ {\begin{array}{*{20}{l}}
	{P\left\{ {i + 1,j - 1\left| {i,j} \right.} \right\} = {2 \mathord{\left/
				{\vphantom {2 j}} \right.
				\kern-\nulldelimiterspace} j},}&{i = 1,2, \cdots ,{L_n} - 1{\rm{; }}j = 2,3, \cdots ,{L_h},}\\
	{P\left\{ {i,j - 1\left| {i,j} \right.} \right\} = {{1 - 2} \mathord{\left/
				{\vphantom {{1 - 2} j}} \right.
				\kern-\nulldelimiterspace} j},}&{i = 1,2, \cdots ,{L_n} - 1{\rm{; }}j = 2,3, \cdots ,{L_h},}\\
	{P\left\{ {i + 1,1\left| {i,j} \right.} \right\} = 1,}&{i = 2, \cdots ,\infty {\rm{; }}j = 1.}
	\end{array}} \right.
\end{align}
\end{footnotesize}
This model is for low network load case, and we can also obtain the modelling for high network load case in the same manner. According to this Markov chain model, it is possible to provide a theoretical guidance for the blockchain implementation, which can help us to analyze the key performance indicators in terms of cumulative weight and confirmation delay under different network loads, and it is able to evaluate the security performance based on the understanding the impact of network loads.
%\subsection{Related Work}
\subsection{Related Work}
\subsubsection{Control of mining difficulty}
\par D. Kraft et al. in  \cite{chapter3RelatedWork1} discuss the difficulty-of-work  re-adjustment in the blockchain system. In order to achieve a relatively ideal average block generation rate over a period of time,   the mining work is formulated as a non-homogeneous Poisson process and  a new method of the difficulty-of-work  re-adjustment is proposed. However, the randomness of the hash rate in the blockchain system has not been addressed yet. To this end, D. Fullmer et al. in \cite{chapter3RelatedWork2} consider this situation and introduces a random model about block arrival time, in which the marginal distribution of block arrival time and its both expectation and variance are derived. Accordingly, we can know that the target difficulty value both is a function related to the arrival time of the previous block and affects the block arrival time in the next retargeting period.

\subsubsection{Modeling and analysis}
\par Using the stochastic reward network (SRN), H. Sukhwani et al. propose a new Hyperledger Fabric v1.0 + system model and study the performance indicators such as throughput, transaction delay, node utilization, and queue length in  \cite{chapter3RelatedWork3}.  In addition, this work analyze the fullsystem as well as the subsystems corresponding to each transaction phase in details.  The proposed model  can provide a quantitative framework that
helps a system architect estimate performance as a function of
different system configurations and makes design trade-offs
decisions.
%In addition, this work provides deep theoretical understanding for blockchain network architecture based on the Fabric, the proposed model and solution method are complex.
N. Papadis et al. in  \cite{chapter3RelatedWork4} propose a stochastic network model to describe the joint dynamics of ``frontier'' processes, track the dynamic evolution of blockchain networks, capture important blockchain features, and study the impact of delay on security.

\subsubsection{blockchain node deployment}
\par  Based on the RAFT consensus mechanism, H. Xu et al. in \cite{8982036} study the security performance of wireless blockchain networks under malicious interference, and provides analysis guidance for the actual deployment of wireless blockchain networks.
  Y. Zhu et al. in  \cite{chapter3RelatedWork5} introduce the blockchain-based heterogeneous network in a air-to-ground IoT heterogeneous network. In order to determine the consensus process and obtain the downlink information transmission rate through mathematical analysis, stochastic geometry method is adopted to model the deployment of Ground Sensors (GSs), Air Sensors (ASs), and the place of eavesdroppers with interference attacks.   Y. Sun et al. in  \cite{chapter3RelatedWork6} model the location deployment and transaction arrival rate in IoT network as a PPP to study the relationship between communication throughput and transaction throughput, and propose an optimal communication node deployment algorithm, which achieves the maximum communication and transaction throughput with the minimum communication node density.   M. Liu et al. in  \cite{chapter3RelatedWork7} formulate the location deployment of base stations and mobile users as a HPPP for mobile edge computing, and derive the theoretical expressions of relevant performance indicators in various modes using stochastic geometric methods.

\subsection{Summary and Existing Problems}
\par Due to the randomness of the blockchain system, stochastic methods is commonly used for formulation, which can accurately describe the distribution of blockchain nodes, the arrival of transactions, the behavior of blockchain users and etc. Through these mathematical modeling,  researchers can track the dynamic evolution of the blockchain system, and further analyze of the consensus process, throughput, and security performance. Accordingly, the corresponding theoretical basis can be provided for performance improving, such as malicious attack preventing, and blockchain application accelerating.
\par Although stochastic process has been widely used in the literature, there are still issues that should be considered and improved in the future.
\begin{enumerate}
  \item Most existing researches formulate the blockchain system as a certain stochastic process assuming a simple model such as the Poisson distribution. However, in the actual environment, the blockchain running process   usually is more complicated, and how to abstract the common random variables accurately without lossing generality should be well studied. In another word, the mathematical formulation must accurately describe the blockchain system in practice, while considering the property, complexity and constraint in the view of theoretical approach.
  \item Currently, some typical issues in a few specific scenarios have been widely investigated. In contrast, a generalized stochastic model is in need to describe the whole blockchain system. In the future, we should further consider how to use stochastic process to model an end-to-end blockchain system model, which can systematical study the actions of blockchain user and the system performance. In addition, understanding the interactions between various functions of hash operation, cache, consensus, communication network and smart contract is another direction for future research.
\end{enumerate}

\section{Game Theory in Blockchain}

\par Game theory \cite{chapter4Introduction1}\cite{chapter4Introduction3} is a mathematical theory to study the strategy selection in competitive behaviors. A basic game consists of four basic elements: player (decision maker), strategy (the player's action), reward (the game result obtained after the player chooses a strategy) and equilibrium (a balance).
\par We can use the game theory to formulate the conflicts and cooperations between selfish and rational decision-makers. By analyzing both expected and actual behaviors of the players, we can study how how each player generate and optimize individual strategy under different situation. In a game, if no player can obtain more profits by changing his own strategy alone, we call that the strategy set of all players at this time is at the state of Nash equilibrium \cite{chapter4Introduction2}. Nash equilibrium guarantees that each player's strategy is optimal no matter how the strategy of other players changes.
\par In recent years, game theory has become an important tool for communication and network research, in which most interactions can be analyzed as game behaviors to find the optimal competitive strategy.

%If we prove that there exists a unique solution in this game, the solution must be Nash equilibrium. Nash Equilibrium belongs to complete information static game: When other players have no incentive to change strategy, the strategy combination is the best choice for all players.

%the user equipment, mobile operator and the MEC server can be regarded as a player, respectively. Due to the different service types lead to different benefits, each player can be motivated in a distributed way to maximize the individual benefit using game theory. Meanwhile, the computing complexity is reduced due to the distributed manner. The limitation of this method lies in the non-uniqueness of Nash Equilibrium, which reduce the effectiveness of Nash equilibrium analysis.\subsection{Model Briefs}
\subsection{Model Briefs}
\par Many blockchain problems can be structured as a game based on game theory. Through the analysis of the optimal strategy and equilibrium solution, we can investigate the operation process of blockchain mechanisms and the behavior of blockchain users to optimize the systematic performance accordingly.
\par Considering the malicious characteristic,   a miner launches an attack to increase his/her own winning probability while unfairly reducing the wining probability of others. In fact, any selfish and rational miner would like to maximize its own overall reward, which is determined by the environment feedbacks, cost and successful attacking probability simultaneously. Therefore, a miner must consider all possible re-actions of others to choose the strategy that is most beneficial to  itself in this typical non-cooperative game. Therefore, the miner can be treated as the player, its strategy is whether to launch an attacking, and the reward function is the expected reward if the attacking succeeds minus the cost for attacking. In this manner, we can formulate this mining process as a game shown in Fig. 8, to study the impact of miners' strategies (to be honest or malicious) on the blockchain network. Based on the analysis and equilibrium solution, a game model could employed and to provide a theoretical guidance to optimize the consensus process in order to improve the security (refrain from launching the attacking action).
% (配个图说明博弈论+安全)
\begin{figure}[H]
  %\centering
  \includegraphics[width=3.5in]{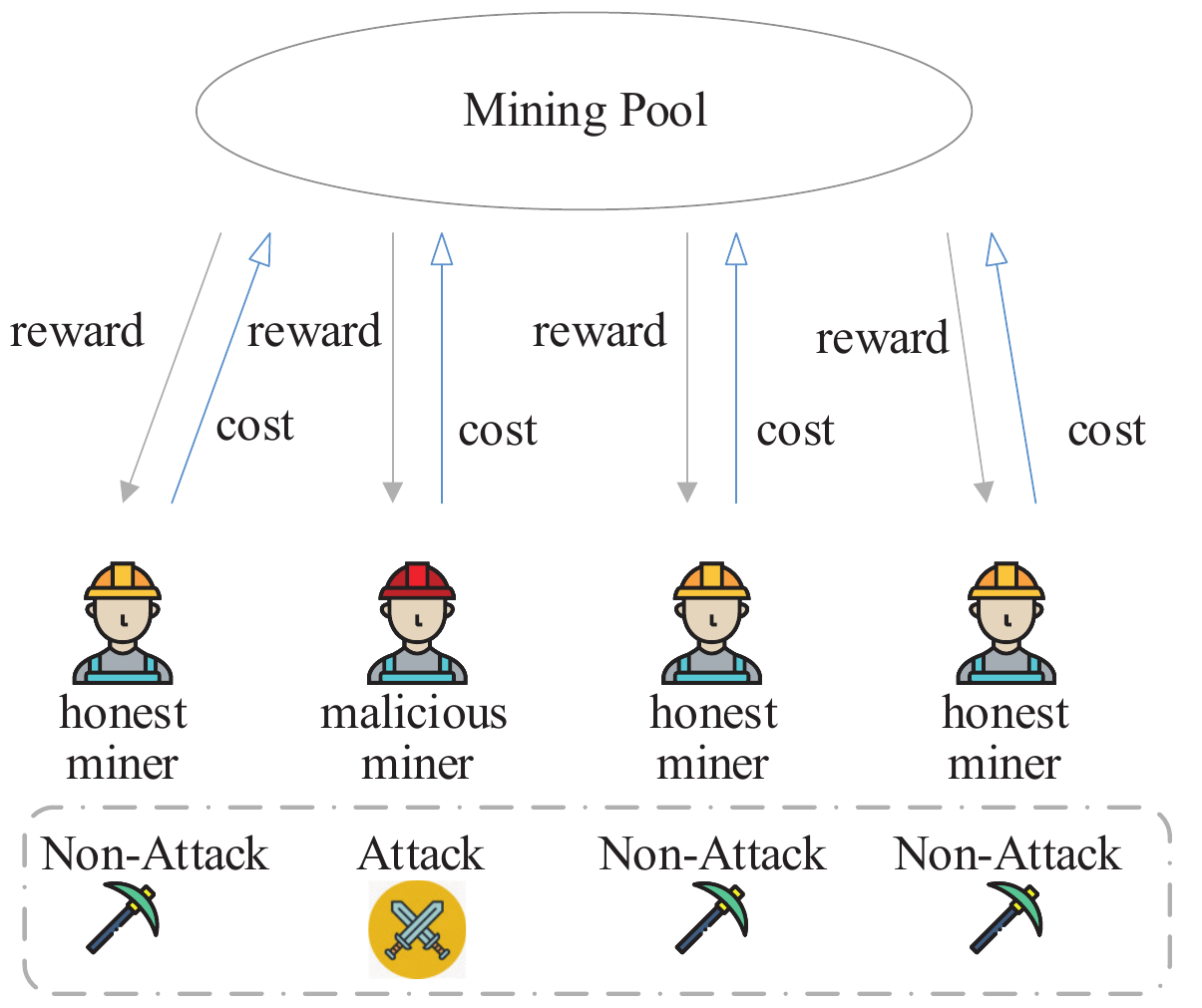}
  % where an .eps filename suffix will be assumed under latex,
  % and a .pdf suffix will be assumed for pdflatex; or what has been declared
  % via \DeclareGraphicsExtensions.
  \caption{\iffalse In Pow-based blockchain network, each miner can choose either attack (Attack) or Non-Attack,  and the miner will get reward according to the strategy. The process can be seen as a game.\fi A Non-cooperative game framework for the mining process }
  \label{fig-4-1}
\end{figure}
\par During the mining process, a miner should allocate certain amount of computing resources   to increase the wining probability to get the right that generates a new block with corresponding reward. Meanwhile, the more computing resources consumed, the higher cost would be generated. It means there is an effective trade-off between cost and benefit needs to be made. Accordingly, this problem can be also formulated as a game to study the interaction among miners and analyze the best strategy. for instance, game theory is often used to study equilibrium-based strategy, which  guarantees the optimal reward of each miner while avoiding the meaningless mining competition caused by greedy resource allocation. Fig. 9 is a game theory model based on auction, the miners will bid for the computing resource, and the computing service providers will provide different computing resource. The more   bid, the more resource. Therefore, the miner need to balance the cost and revenue.
%(配个图说明博弈论、auction)
\begin{figure}[H]
  %\centering
  \includegraphics[width=3.5in]{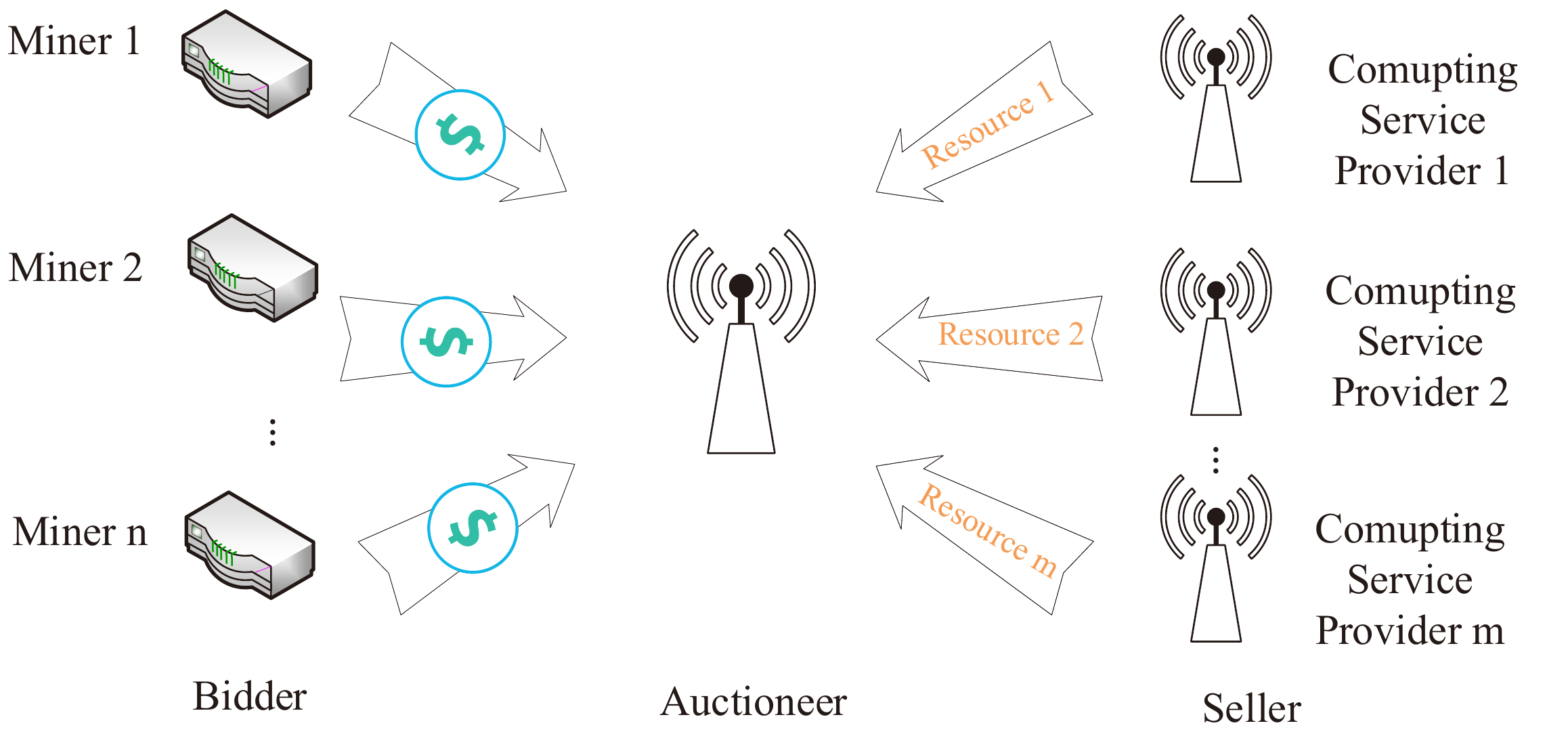}
  % where an .eps filename suffix will be assumed under latex,
  % and a .pdf suffix will be assumed for pdflatex; or what has been declared
  % via \DeclareGraphicsExtensions.
  \caption{ A auction-based a non-cooperative game model  for computing  resource allocation.  }
  \label{fig-4-2}
\end{figure}
\par Using mobile edge computing in blockchain networks, the miner can offload its mining task to edge server, which can solve the limitation of computing resource of blockchain users to extend the blockchain application in wireless scenarios effectively. To encourage the miner to offload reasonably and motivate edge server to process effectively, the miner should pay an amount of payment to edge server for offloading. As shown in Fig. 10, the miners will offload some mining task to mobile edge. The more computing resource the miner buy, the greater the probability of   successfully minings. Stackelberg game can be employed based on the concerning of resource allocation as mentioned above. If addressing the mining task assignment (the association between edge server and miner), auction model is a common approach.
% (配个图说明博弈论+mec)

\begin{figure}[htb]
  %\centering
  \includegraphics[width=3.5in]{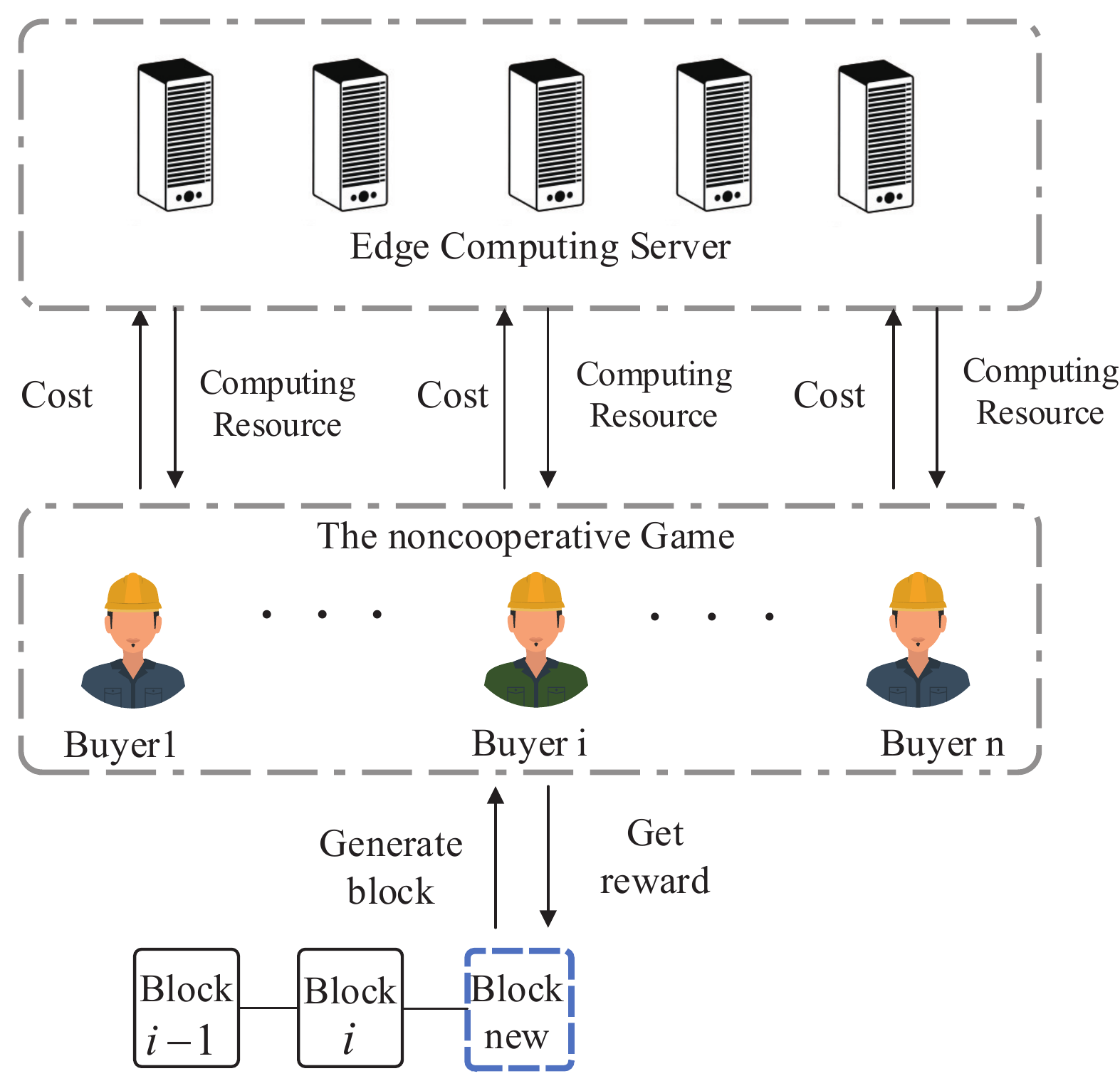}
  % where an .eps filename suffix will be assumed under latex,
  % and a .pdf suffix will be assumed for pdflatex; or what has been declared
  % via \DeclareGraphicsExtensions.
  \caption{A MEC-basd non-cooperative game  for mining strategy}
  \label{fig-4-3}
\end{figure}
\subsection{Case Study}
\par In this section, we will use our previous work \cite{chapter4CaseStudy} as an example to introduce how to motivate honest   actions of participants in blockchain-based scenarios.
\par In the typical MEC enabled WBN, controlled by the MEC manager,
edge servers are just treated as network resources providers,
including computational resources and storage resources. However,
the MEC server manager may become a central node that is independent of the blockchain system,
thereby undermining the distributed nature of the blockchain system and further damaging its security.
To this end, in this example, the underlying P2P network consists of edge servers,
who are treated as blockchain miners,
undertaking blockchain functionality operations and earning transaction fees,
while IoT devices are treated as blockchain users,
who only need to upload transactions to blockchain
with specified transaction rate requirements. As shown in Fig. 11,  the blockchain users submit  transactions to  blockchain miners, and then the blockchain miners execute  the mining task and successfully generate   a block. Finally, it will be added to the local ledger and broadcast to peers and this operations will also be performed by other blockchain miners once they have verified it as a valid block.

\par At blockchain users' side, the utility of a blockchain user $f_{j}$ includes the satisfaction degree and incentive cost, i.e., the transaction fee.
Thus, in order to maximize the utility by requiring considerable transaction rate $\gamma_{j}$, the optimization problem for $f_{j}$  can be expressed as follows:
\begin{align}
\begin{split}
\max \limits_{ \gamma_{j}}\quad &U_{f_{j}} = S_{f_{j}}(\gamma_{j} )-C_{f_{j}}(\gamma_{j} )\\
&s.t. \sum_{j=1}^{N}\gamma_{j}\leq \Gamma_{max}
\end{split}
\end{align}
where $S_{f_j}(\gamma_j)$  is the satisfaction degree, $C_{f_j}(\gamma_j)$ is the transaction fees and $\Gamma_{max}$ is the maximum transaction rate blockchain system can afford.
\par At the blockchain miners' side, the utility of them is defined as charged transaction
fees minus computational resources consumption.
Thus, to maximize their revenue, the optimization problem can be expressed as
\begin{align}
\max \limits_{ \beta} \quad U_{l} = S_{l}(\sum_{j=1}^{N}\gamma_{j},\beta)-C_{l}(\sum_{j=1}^{N}\gamma_{j})
\end{align}
where $N$ is the number of  blockchain users, $ S_{l}(\sum_{j=1}^{N}\gamma_{j},\beta)$ is the earning by publishing $\sum_{j=1}^{N}\gamma_{j}$ transactions per hour
and $ C_{l}(\sum_{j=1}^{N}\gamma_{j})$ is the corresponding cost of resources consumption.

\begin{figure}[htb]
  \centering
  \includegraphics[width=3.5in]{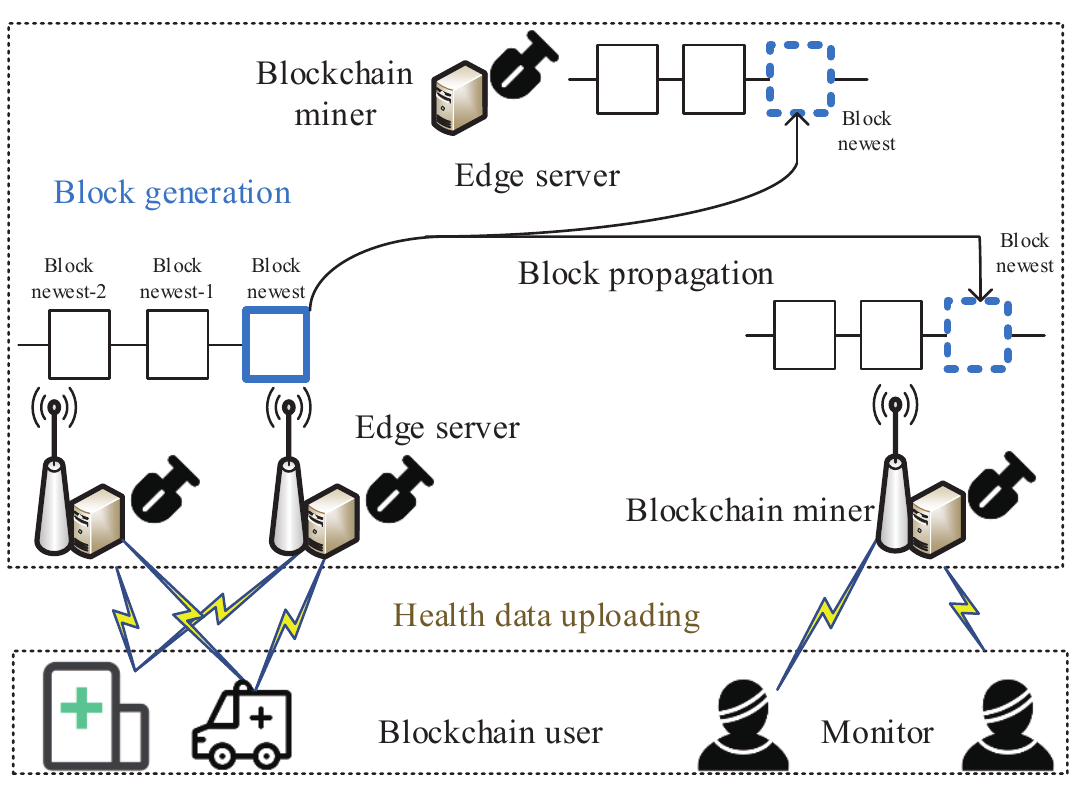}
  % where an .eps filename suffix will be assumed under latex,
  % and a .pdf suffix will be assumed for pdflatex; or what has been declared
  % via \DeclareGraphicsExtensions.
  \caption{The explaination of the case study in Game Theory \cite{chapter4CaseStudy}}
  \label{fig-4-CaseStudy}
\end{figure}

\par With blockchain miners acting as the leader while blockchain
users acting as followers,
a single-leader-multiple-followers Stackelberg game
can be used to model interaction between them.
Based on the Karush-Kuhn-Tucker (KKT) conditions
and backward induction method, a distributed algorithm
can be designed to reach the optimal strategy $(\beta^*,\gamma_j^* ) $
in an iterative manner.

\subsection{Related Work}

%\subsection{Future Work}
\subsubsection{Mining pool management}
\par Game theory has been widely used for mining pool management. In  \cite{chapter4RelatedWork1}, X. Liu et al. use an evolutionary game to describe the dynamic evolution of pool selection strategies for individual miners, analyze the evolution of the mining pool selection strategy considering the hash rate, and the broadcast delay of the block. In  \cite{chapter4RelatedWork2}, J. Li et al. define the mining as a non-priority queuing problem that is determined entirely by transaction fee, and propose a transaction queuing game model to study the role of transaction fee in the consensus process. Based on this model, the authors analyze the relationship between the mining reward and the time cost, and prove the existence of Nash equilibriums. In order to improve the mining rate, the authors in  \cite{chapter4RelatedWork3} and  \cite{chapter4RelatedWork4} formulate the mining process in a PoW-based blockchain as iterative game, and apply Zero-determinant strategy to optimize the mining strategy in order to solve the miners' dilemma problem.

\subsubsection{Security strategy}
\par Game theory is also potential to study the behaviors of blockchain users and the strategies for security concerns. S. Feng et al. in \cite{chapter4RelatedWork5} introduce a risk management framework for blockchain service, and a Stackelberg game is adopted to describe the interactions among blockchain providers, network insurance companies and blockchain users. Based on this game model, the existence and uniqueness of equilibrium are discussed, and the three-party equilibrium-based strategy is analyzed to avoid the double-spending attacking. S. Kim et al. in \cite{chapter4RelatedWork6} use the evolutionary game to study the dynamics of mining pool strategy, in which the pool can choose some participating miners to infiltrate into other pools to launch a block withholding attack. Based on the formulated model, the authors qualitatively analyzed the influence of malicious infiltrators on mining pool strategy and the feasibility of automatic migration among pools.

\subsubsection{Resource management}
\par In order to study the issues of resource management and pricing between cloud computing providers and miners, H. Yao et al. in   \cite{chapter4RelatedWork7} propose a multi-agent reinforcement learning algorithm to find the Nash equilibrium of the proposed model, and prove that the Nash equilibrium point of service demand in the system is related to the expected reward of each miner. Y. Jiao et al. in \cite{chapter4RelatedWork8} propose  an auction-based model to study the interaction between miners and edge service providers, and analyze the allocation and pricing of edge computing resource in the blockchain network. Considering the reward of service providers, N. C. Luong et al. in  \cite{chapter4RelatedWork9} propose an optimal auction model using deep learning to solve service providers reward and resource management issues. D. Xu et al. in  \cite{chapter4RelatedWork10} study the security issues in blockchain edge networks using the game theory as well. In this work, a penalty scheme based on behavioral records is designed considering the conditions of Nash equilibrium.

\subsection{Summary and Existing Problems}
\par As an analysis tool, game theory is widely used to study security, mining, and resource allocation problems in blockchain networks. A game model can be built by capturing the characteristic of the addressed problem  in terms of the role of blockchain users, behavior of blockchain decision-maker and the performance of blockchain system. Thus, the game modling can help researchers understand the impact of different strategy and analyze the optimal strategy based on equilibrium solution.
%Capturing the characteristic of the addressed problem to describe the role of blockchain users, behavior of blockchain decision-maker and the performance of blockchain system as a game model, it is helpful to understand the interaction of different strategy and analyze the optimal strategy based on equilibrium solution.
\par Meanwhile, some problems are still remaining to be addressed as follows.

\begin{enumerate}

  \item Some necessary information should be collected and exchanged in game theory, for example, the selling price of edge server for mining task offloading and the computing resources used by miners for mining. Schedule the communication resource into the game theory need to be well investigated.
  \item Furthermore, to achieve the equilibrium solution, multi-round iterations in game theory are needed usually, but the corresponding overhead should be considered. The overhead which cause commnuication delay, would generate significant impact on the performance and the security of blockchain system.  Next, efficient and light-weight game theoretic mechanism for blockchain system should be investigated.
  \item Rationality and selfishness are the basic assumptions in game theory. However, these assumptions might be invalid especially for the malicious attacker, whose purpose is to launch an attacking to ruin the blockchain system regardless of the cost. Therefore, how to understand and study this extreme malicious action in order to improve security and privacy should be studied in the future.
\end{enumerate}

\section{Optimization in Blockchain}

Optimization theory \cite{chapter5Introduction1}\cite{chapter5Introduction2}  is a well-known computational tool to solve theoretical analysis and practical engineering issues. Generally, the goal of the optimization theory is to make the best decisions under some constraints. As a famous subfield of optimization theory, convex optimization \cite{chapter5Introduction1} has been widely investigated and applied. Since lots of optimization problems can be transformed into a convex optimization, this section mainly focuses on the applications of convex optimization in blockchain. In addition, because optimization problems in large-scale and complicated network environments are usually nonconvex, convex optimization can also be effectively applied by relaxing or/and approximating some nonconvex conditions. A basic form of convex optimization can be regarded as constrained optimization problems which can be formulated as
\begin{align}
\begin{split}
\begin{array}{l}
	\mathop {\min }\limits_x f\left( x \right)\\
	s.t.\left\{ {\begin{array}{*{20}{c}}
{{g_i}\left( x \right) \le 0,{\rm{ }}i = 1, \cdots ,m}\\
	{a_j^T = {b_j},{\rm{ }}j = 1, \cdots ,p}
	\end{array}} \right.
\end{array}
\end{split}
\end{align}
where the objective function $f(x)$ and the inequality constrained function $g(x)$ are convex functions on ${{\mathop{\rm R}\nolimits} ^n}$, and the equality constraint function ${h_i}\left( x \right) = a_i^Tx - {b_i}$ must be affine.

%这里还缺乏内容

\subsection{Model Briefs}

\par Optimization theory is widely used in various fields, which help us to adjust the parameters, schedule the resource and determine the decision improving the system performance in a distributed/centralized manner. In blockchain, it can be used to manage the mining, improve security, and jointly optimize the blockchain with other technologies.
The mining management optimization is a typical problem in blockchain. As well known, the mining reward for miners is an important factor to encourage the contribution of computing resource. Therefore, maximizing the reward and promoting the accomplish of the consensus process is a popular topic in mining management. In order to increase the successful mining probability, miners usually choose to join the mining pool. Generally, as long as any miner in the pool succeeds in mining, the mining reward will be distributed to each miner in the pool. Different mining pools may adopt different reward mechanisms. Therefore, the miner should consider how to choose the optimal pool selection strategy to optimize its reward, and this problem can be transformed into a mathematical optimization problem. For this optimization problem, the mining reward obtained by miners under different reward mechanisms can be formulated as the objective function, the variable is the miner's pool selection strategy, and the constraints are   determined by the actual situation. Through this mining pool selection optimization problem, we can study the impact of different reward mechanisms on the blockchain network in term of the objective function, and determine the optimal selection under the constraints.
% (配图说明mining中的优化)
%note, 由于不好配图,取消
\par When miners join the mining pool and cooperate with others, the competition among miners becomes the competition among the mining pools. In order to win for mining reward, some mining pools may choose to take the Block Withholding (BWH) attacking on other mining pools. As shown in Fig. 12,  the pool A will allocate a part of computing power $\alpha$ to another pool B for attacking.  The greater computing power consumed by the attacker, the greater the malicious impact on other pools would be happened.  The malicious impacts would decline the computing power for consensus and is costly to the attacker itself. As a result, the attacker should choose the optimal computing power for attacking, which can be considered as an optimization problem. In this case, the objective function is to maximize the mining reward as well as successful attacking probability with the computing power for attacking as the variable. Through the optimization analysis, we can know the optimal attacking strategy, which is the baseline to analyze and design the consensus process for security based on the understanding the malicious action of the attacker.
%(配图说明优化+安全)
\begin{figure}[]
  \centering
  \includegraphics[width=3.5in]{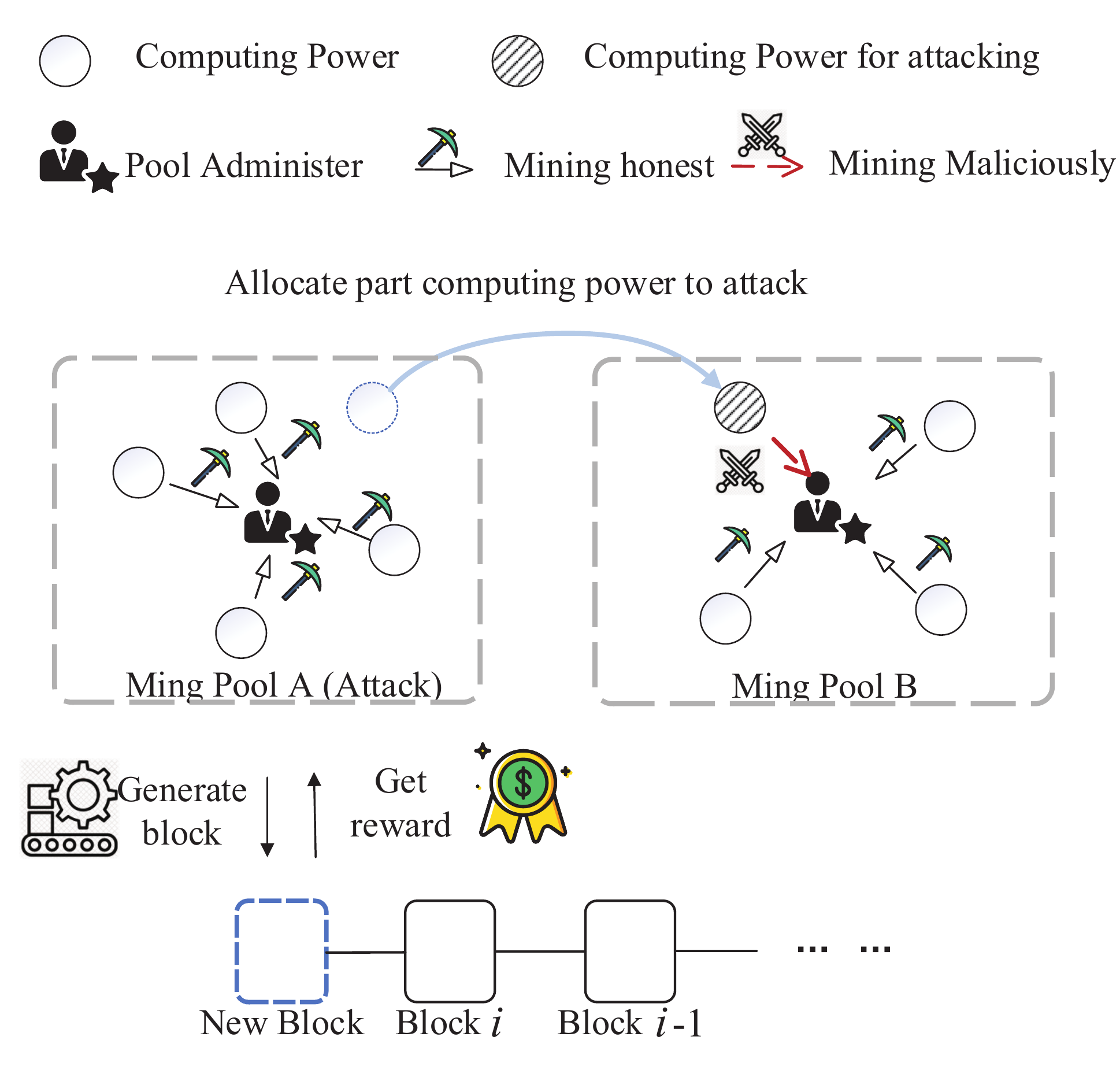}
  % where an .eps filename suffix will be assumed under latex,
  % and a .pdf suffix will be assumed for pdflatex; or what has been declared
  % via \DeclareGraphicsExtensions.
  \caption{The explaination of  BWH attack.}
  \label{fig-5-2}
\end{figure}
\par As discussed before, mining task offloading from the miners to the edge servers can be also formulated as an optimization problem with resource allocation. In this case, the miner should choose an appropriate offloading strategy to determine whether to offload, or to which edge server to offload, and how many resources (computing, communication and cache) the edge server allocates to the offloading task. Using this optimal offloading formulation, we can analyze the impact of the offloading on the performance and security of the blockchain network, and the optimal solution is to provide a theoretical guidance for the blockchain development and application.
%（配图）删掉了

\subsection{Case Study}
\par In this section, we use the previous work  \cite{chapter5CaseStudy}   as an example to introduce how to perform optimization in blockchain.
In this paper, to improve the security and privacy of D2D (device to device) communication, the authors introduce a new distributed and secure data sharing framework called D2D blockchain. The simplified procedure for transaction relaying and  block verification  is shown in Fig. 13. In this framework, the authors deploy a series of Access Point (APs) to incentivize the transaction relay and block verification using DPoS-Based Lightweight Block Verification Scheme. Therefore, an important problem for AP is that how to improve the efficiency of transaction relaying and DPoS based block verification, while the payment (cost) is small.

%（配图）
\begin{figure}[ht]
  \centering
  \includegraphics[width=3.5in]{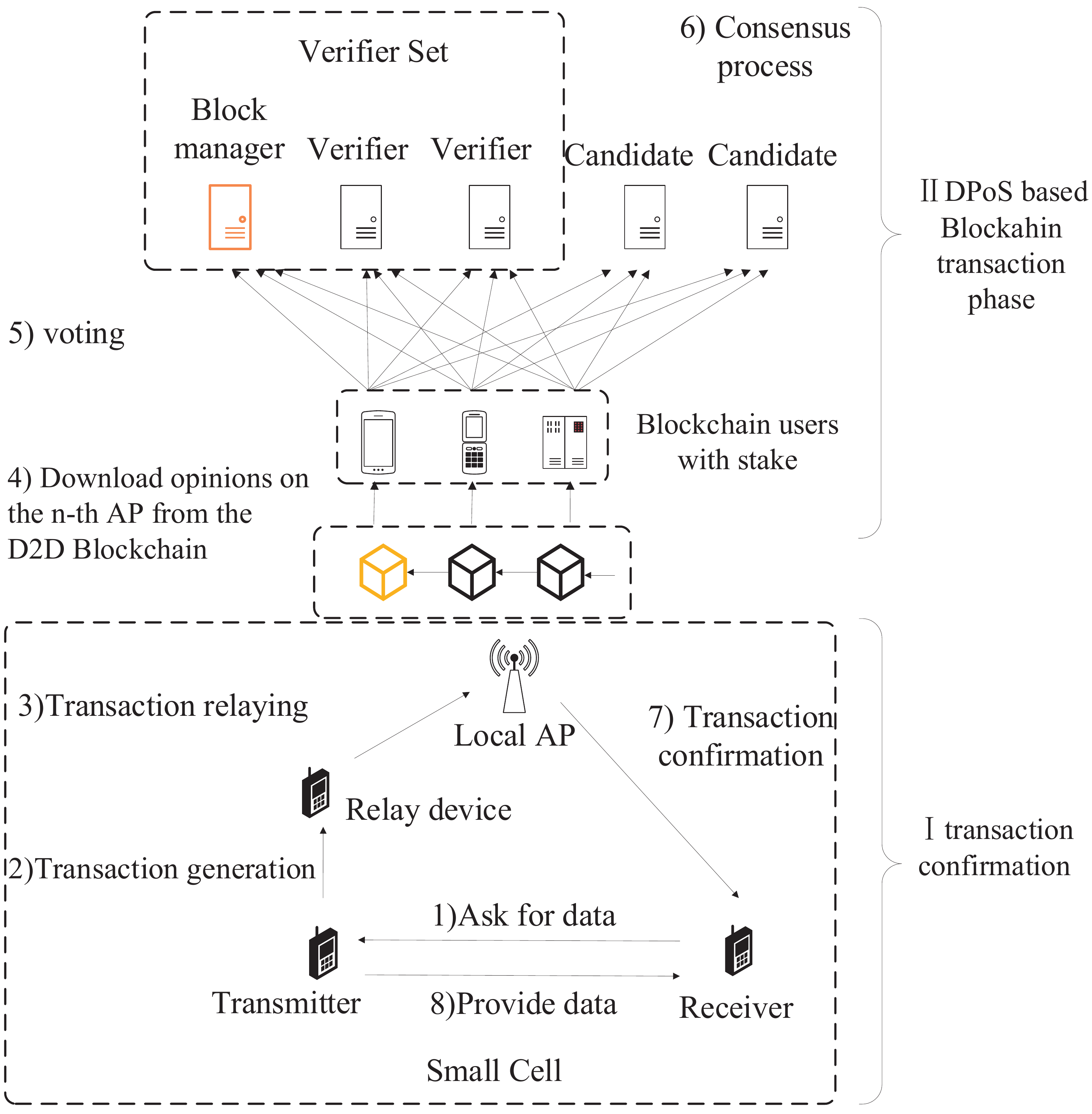}
  % where an .eps filename suffix will be assumed under latex,
  % and a .pdf suffix will be assumed for pdflatex; or what has been declared
  % via \DeclareGraphicsExtensions.
  \caption{The simplified procedure for transaction relaying and  block verification \cite{chapter5CaseStudy}}
  \label{fig-5-CaseStudy}
\end{figure}

\par In this work, a two-stage contract theory based on joint optimization scheme is proposed, where the AP serves as an employer who designs all kinds of contracts, pays rewards for employees, relays devices, and verifiers serve as employees. Relay devices should consider their battery energy, resource of occupied bandwidth, etc., and verifiers should consider their CPU cycles, energy consumption, etc. In order to maximize the expected utility of AP while satisfying the individual rationality and the incentive compatible constraints for transaction relaying and block verification, the objective function in the optimization problem is formulated as
\begin{align}
\begin{split}
   max \quad &U_{AP}=V_R-R_R+V_V-R_V \\
& s.t. \quad \{a,b,c,d,e\}
\end{split}
\end{align}
where the  limiting conditions $ \{a,b,c,d,e\}$ are respectively expressed as

\begin{itemize}

  \item[(a)] For each relay device, the reward paid from AP should be not less than its cost due to the rationality.
  \item[(b)]  Similarly, for each verifier, the reward paid from AP should be not less than its cost due to the rationality.
  \item[(c)]  AP needs to design the contract for the relay device or verifier flexibly according to their corresponding types.
  \item[(d)]  Similarly, AP needs design the contract for the verifier flexibly according to their corresponding types.
  \item[(e)]  For AP, the reward paid to relay devices and verifiers must be reasonable following a limitation.
\end{itemize}
where $V_R$ is the whole value of transaction relaying created by all the relay devices, and $R_R$  is the whole payment from AP to relay devices. Similarly, $V_V$  is the whole value of block verification created by all the verifiers, and $R_V$ is the whole payment from AP to verifiers.
\par However,  to solve the optimization problem in a practical system, the constraint conditions often are  set to  non-convex, thus (5.2) cannot be solved directly.

\par By reducing some constraints, this optimization problem can be transferred into a convex problem. Finally, the optimal solution is the best strategy for transaction relaying and block verification, which can maximize the utility of AP while incentivizing the relay devices and block verifiers to accomplish their tasks optimally.

\subsection{Related Work}
\subsubsection{Security}
\par As one of the most important issue in blockchain, the security topic has been widely formulated as an optimization problem. O. Onireti et al. in   \cite{9013778} propose a practical modeling framework for PBFT, and the viable area for wireless PBFT network is defined to ensure the minimum number of replication nodes required for protocol security and activity. Considering the secured communication and data sharing between vehicles, J. Kang et al. in  \cite{chapter5RelatedWork1} propose a two-stage security enhancement solution to solve collusion attacks in  IoV. In the first stage, the system selects active miners and standby miners (candidates) based on reputation voting. Hence, the active miners selected get the chance to generate block. In the second stage, standby miners   will verify the block generated by active miners. Therefore, the internal collusion among active miners is avoided. However, how to incentivize the standby miners to participate is  an important problem, which is solved using contract theory. M. Saad et al. in  \cite{chapter5RelatedWork2} model the malicious behavior as a Lyapunov optimization problem. Then they study how attacker uses the impact of memory pool overflow on blockchain users to launch an DDoS attacking. To prevent DDoS attacks, the authors propose two effective countermeasures: fee-based and age-based design. The core of the fee-based design is to conduct transaction relay with minimum relay fee to reject spam transactions. And similarly, the core of the age-based design is to compare transaction mining fee and minimum mining fee. While spam transactions are rejected, the DDoS attacks are also resolved.

\subsubsection{Resource allocation}
\par Considering the high-resource consumption, mobile edge computing is a nature design for the blockchain-enabled wireless networks.   Y. Wu et al. in \cite{chapter5RelatedWork3} convert the computing resource allocation problem in  multi-access MEC-based blockchain into a mathematical form of joint optimization problem. To maximize the total revenue of the mobile terminals while ensuring the fairness of the mobile terminals, they consider two different scenarios, namely a single-edge-server scenario and a multi-edge-servers scenario. For the two scenarios, the authors propose two layered algorithms to solve the non-convex optimization problem above. S. Fu et al. in  \cite{chapter5RelatedWork4} study the issue of joint resource allocation in blockchain-based IoT systems. To maximize the system energy-efficiency, the authors use stochastic programming to solve the joint optimization problem above.
%新增随机规划方法的文献
M. Wang et al. in  \cite{chapter5RelatedWorAdd1} propose a blockchain-based decentralized and truthful framework for MDC (BC-MDC), which enable the decentralization and prevented dishonesty by incorporating a plasma-based blockchain into the MDC. There are 4 smart contracts designed for distributed management of worker registration, task posting/allocation, rewards and penalties. Furthermore, MDC task allocation is formulated as a stochastic optimization problem that jointly minimized the long-term processing cost and risk of task failure, and an online task allocation is proposed to timely allocate   the task to workers to accommodate the network variation. Moreover, a rewarding/penalizing scheme is designed to ensure the individually rational to truthful of rewarding and prevent the free-riding as well.

\subsubsection{Optimal algorithm and strategy design}
\par
The optimization problem can be also used to describe various purposes in blockchain system.   Y. Zhang et al. in  \cite{chapter5RelatedWork5} study the routing issue in a blockchain-based payment channel network. The authors analyze the payment routing problem using the optimization theory method. While considering the constraints of timeliness and feasibility, the authors propose a distributed optimization scheme to achieve the lowest total transaction costs from the sender to the receiver. Applying the consortium blockchain technology to electric taxi charging scenarios with multiple operators, J. Zhang et al. in  \cite{chapter5RelatedWork6} propose a new Byzantine fault tolerance algorithm to solve the problem of trust among operators of charging stations. In addition, this work designs a system model based on multi-objective optimization to maximize operating efficiency and customer satisfaction while minimizing the time and distance costs of electric taxis.
%随机规划新增文献
Z. Jin et al. in  \cite{chapter5RelatedWorAdd2} propose EdgeChain, a blockchain-based architecture to make mobile edge application placement decisions for multiple service providers. The placement decision is modeled as a stochastic programming problem to minimize the placement cost for mobile edge application placement scenarios. In this scenario, the blockchain is used to store all placement transactions, which can be traceable by every mobile edge service providers and application vendors who consume resources at the mobile edge.
\subsection{Summary and Existing Problem}
\par For blockchain system, the mathematical optimization tool is usually introduced to find the best pool selection strategy for miners to determine the best task offloading strategy, and to allocate resource for consensus process. Although existing researches show that the optimization can achieve the better system performance and security, some problems should be further studied in the future.
\begin{enumerate}
  \item The optimization problem and corresponding solution are usually for a specific situation in static or semi-static state. However, the practical blockchain system is dynamic and stochastic with some important parameters and constraints that are uncertain and might be changed over time and space.  Therefore, these uncertainties should be further considered in optimization formulation.
  \item Due to the complexity of the blockchain system, the solution to  the formulated optimization problem might be challenging. For example, the original problem should be transformed into a standard convex problem, decompose into multi-subproblem, or design an iterative approach to achieve a sub-optimal solution instead of the optimal one. Therefore, the solution method is another issue that should be addressed considering the solution quality, convergence speed and overhead.
  \item For modelling, some basic assumptions and simplifications with some typical mathematical properties are necessary especially for problem formulation and analysis, but they cannot describe the actual blockchain system accurately. Therefore, the acceptable gap between the reality (practical system) and ideality (mathematical model) should be also well studied.
\end{enumerate}
\section{Machine Learning in Blockchain}

\par Machine learning \cite{chapter6Introduction5}\cite{chapter6Introduction6} is a method of designing and analyzing algorithm to ``learn" automatically, which allows computers to analyze from a large amount of   data, find out the hidden laws for prediction or classification based on characteristics of  data.
 Machine learning usually involves  the algorithm of supervised learning \cite{chapter6Introduction3}, unsupervised learning, and reinforcement  learning \cite{chapter6Introduction2}, and the model of Surport Vector Machine (SVM) \cite{8919978}, Random Forest (RF) \cite{1911.11592}, and Deep Learning (DL) \cite{chapter6Introduction1}\cite{chapter6Introduction4}. Those models and algorithms mentioned above   are widely used in the analysis, prediction, and optimization of communications and networks.
Supervised learning is to train labeled data and analyze the training data to solve classification and regression problems. In contrast, unsupervised learning trains data with no labels to achieve clustering or dimensionality reduction by finding similarities or internal relationships in the data. Different from supervised/unsupervised learning, reinforcement learning is mainly used to solve decision-making problems in a trial-and-error process based on the interaction and feedback between the agent and environment.
%Particularly, the machine learning models of SVM and RF are based on supervised learning.
%SVM is a reliable model that does not rely on a large amount of data, which applies to problems where the features are clear but there is no huge amount of data.
SVM discriminates two classes by fitting an optimal linear separating hyperplane to the training samples of two classes in a multidimensional feature space \cite{4358858}.
Random forest is a more accurate and stable model obtained by building multiple decision trees and fusing them together, with the advantage of high accuracy and efficient operation \cite{randomforests}, which is suitable for the case of non-differentiable model with discrete features and limited values.
Deep learning allows computational models that are composed of multiple processing layers to learn representations of data with multiple levels of abstraction \cite{DL}.
It involves  AtuoEncode, Variational Auto-Encoder and Generative Adversarial Network based on unsupervised learning, Deep Neural Networks, Convolutional Neural Networks (CNN), and Recurrent Neural Networks (RNN) based on supervised learning \cite{DL1,DL2}. It is based on basic features and uses multi-layer activation functions to learn high-dimensional nonlinear features, including CNN networks suitable for image domains, RNN networks suitable for time series and WiDE $\&$ Deep networks suitable for recommendation domains, etc \cite{wideanddeep}.

\subsection{Model Briefs}

\par In recent years, machine learning has been widely used in pattern recognition \cite{patternrecognition}, data mining \cite{datamining}, etc., due to its capabilities in data management, analysis, and decision-making. In blockchain, machine learning can provide an efficient and intelligent approach to discover the malicious action and recognize the attacks to guarantee the data reliability, system security and user privacy. On the other hand, machine learning is also used to make the decision for resource allocation, block size setting and transactions scheduling to optimize the performance of blockchain system.
\par Malicious attackers can launch double spend attacking \cite{doublespend}, denial of service attacking, and eclipse attacking on the blockchain network, which will cause the deteriorated security risk. In recent years, machine learning is introduced in blockchain to solve security problems. By using unsupervised/supervised learning algorithms such as the K-means algorithm and  supervised  SVM, etc., we can monitor the transaction in consensus process and the behavior of blockchain users  \cite{chapter6RelatedWork1}. Accordingly,  as shown in Fig. 14, we can learn the characteristic of both  honest and malicious actions,   identify suspicious transaction, and malicious attacker or illegal activity in the network, which can reduce the possibility of successful attacks.
\begin{figure}[H]
  \centering
  \includegraphics[width=3in]{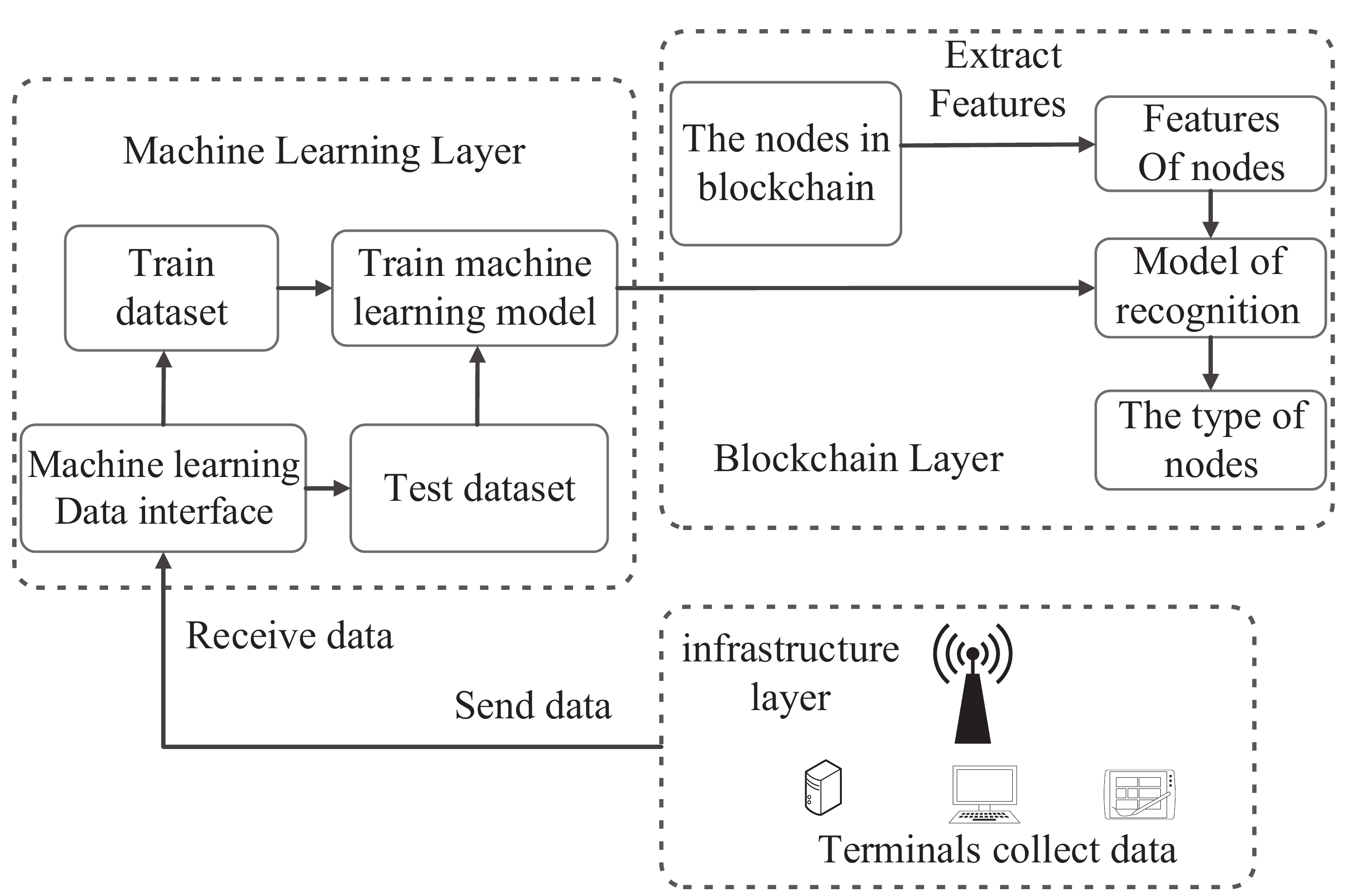}
  % where an .eps filename suffix will be assumed under latex,
  % and a .pdf suffix will be assumed for pdflatex; or what has been declared
  % via \DeclareGraphicsExtensions.
  \caption{Using machine learning to recognize the type of blockchain nodes.}
  \label{fig-6-1}
\end{figure}
\par Nowadays, the resource consumption is an   barrier for blockchain applications, especially in the case which is resource-limited such as the IoT device in the wireless network \cite{2003.13083}. Therefore,
it is necessary to consider the energy-saving in mechanism design and resource allocation. To develop an optimal strategy, we can define the state space $s(t)$, the action space $a(t)$, and the reward function $r(t)$ to represent the agent, environment, and the feedback between them in a machine learning manner \cite{reinforcelearning}. The interaction among $s(t)$, $a(t)$ and  $r(t)$ is shown in Fig. 15. In a real environmental transformation, the probability of going to the next state $s(t+1)$, is related to both the current state $s(t)$ and the previous state $s(t-1)$, occasionally related to even earlier state $s(t-2)$, so that the environmental transformation model may be  too complex to model. Therefore, the way to simplify the environmental transformation model of reinforcement learning is to assume the markov property of state transformation: the probability of transformation to the next state $s(t+1)$ is only related to the current state $s(t)$, and has nothing to do with the previous state \cite{reinforcementlearningmarcov}.
 Accordingly, to maximize the long-term reward, the optimal strategy determination considering the interaction and feedback over time can be treated as a Markov decision process, and it is able to be achieved using reinforcement learning or deep reinforcement learning algorithms.
%%插图
\begin{figure}[H]
  \centering
  \includegraphics[width=3in]{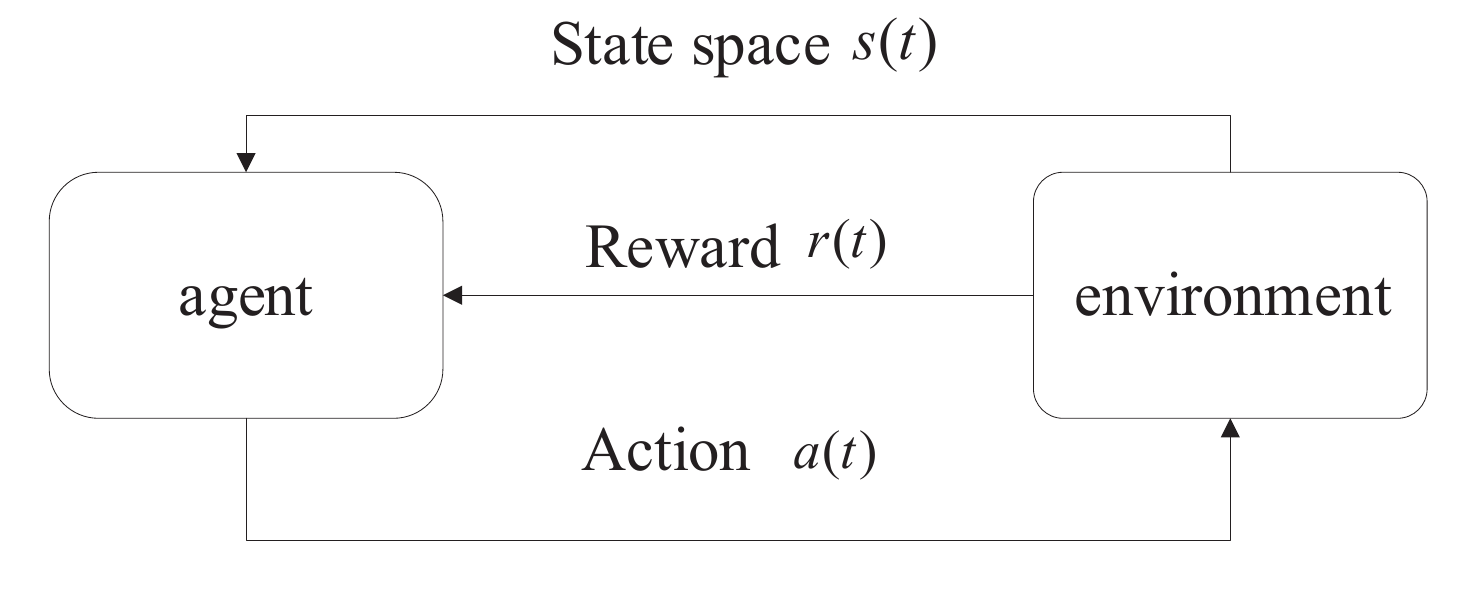}
  % where an .eps filename suffix will be assumed under latex,
  % and a .pdf suffix will be assumed for pdflatex; or what has been declared
  % via \DeclareGraphicsExtensions.
  \caption{The explaination of the  interaction and feedback between  agent and environment  in a machine learning manner.}
  \label{fig-6-2}
\end{figure}
\par In addition, blockchain is a potential solution to the problems   in machine learning. Most typically, the decentralized feature of blockchain can be used to solve the problem of single point of failure caused by aggregating machine learning models using centralized servers.
\subsection{Case Study }

\par In this section, the previous work \cite{chapter6CaseStudy}
is used as an example to illustrate the benefits
of the combination of blockchain and machine learning.
\par
\begin{figure}[h]
  %\centering
  \includegraphics[width=3.5in]{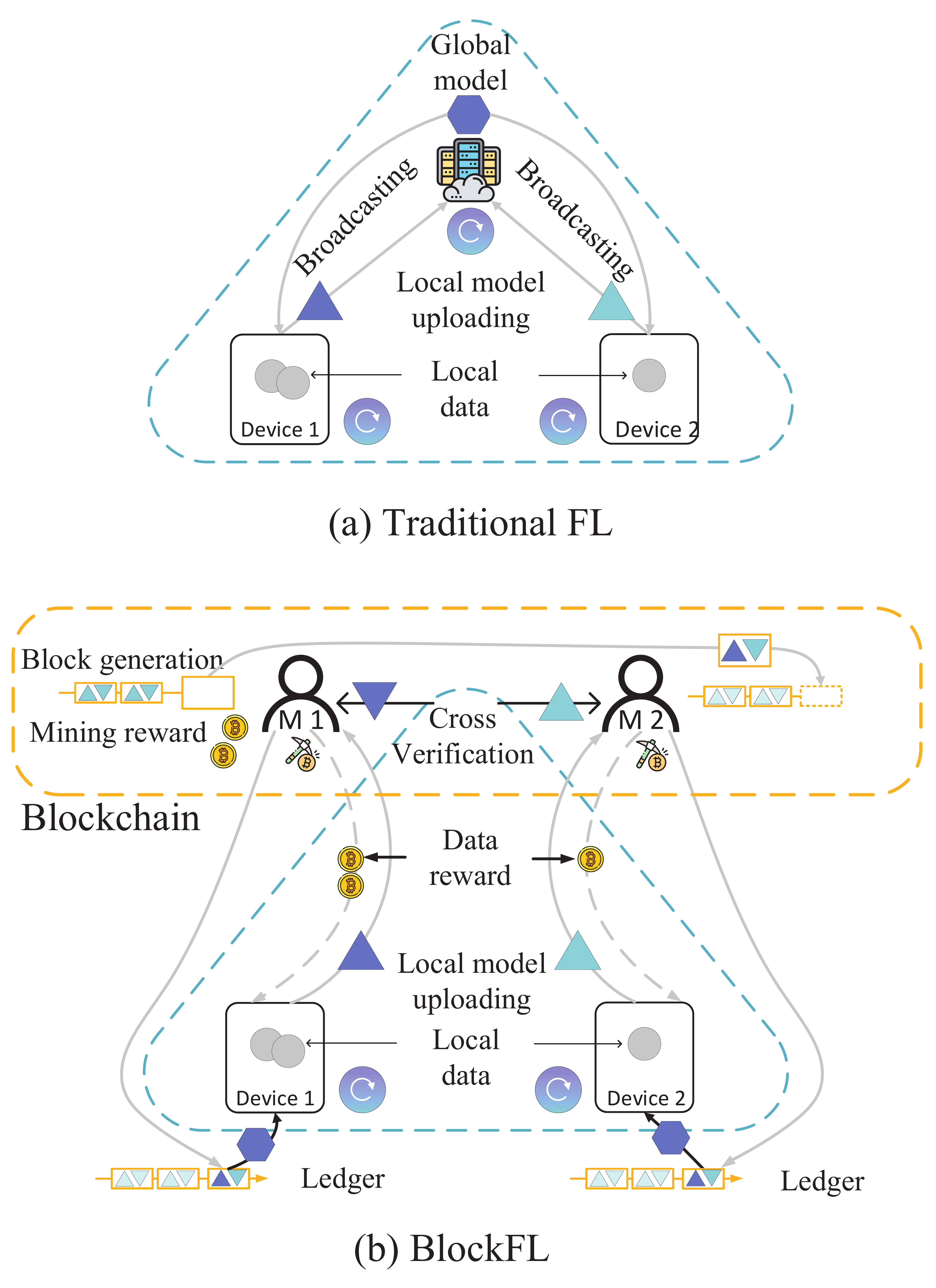}
  % where an .eps filename suffix will be assumed under latex,
  % and a .pdf suffix will be assumed for pdflatex; or what has been declared
  % via \DeclareGraphicsExtensions.
  \caption{ (a) The structure of traditional FL; (b) The structure of the proposed BlockFL \cite{chapter6CaseStudy} }
  \label{fig-6-CaseStudy}
\end{figure}

 Federated learning (FL)
 \cite{chapter6CaseStudy2},
as a promising training paradigm,
was proposed by Google to tackle the privacy
and security problems of centralized machine learning
and to alleviate the communication load of the core network.
As shown in Fig. \ref{fig-6-CaseStudy} (a), many devices collaborate
in solving a machine learning problem
by updating the local model with their own local data,
under the coordination of the centralized FL server.
However, the traditional FL, which relies
on a centralized FL server for model aggregation
is vulnerable  to experience service paralysis even when a single point of failure happens.
All local models updated from devices will be distorted
by the inaccurate global model aggregated
at the FL server.
In addition, for some devices with massive amounts of data, if there is no credible incentive mechanism, they are usually unwilling to participate in training, which brings great challenge  to the rapid convergence of the FL model.
%In addition, there is a great challenge for the rapid convergence of the FL model due to the lack of credible incentive mechanisms to devices that own massive data and have no willingness to participate in training.

\par To address the problems mentioned above,
H. Kim proposed a blockchained FL (BlockFL) architecture,
which is shown in Fig. \ref{fig-6-CaseStudy}(b). With the distributed and non-tamperable characteristics of the blockchain, the use of a blockchain network instead of the centralized FL server can effectively overcome the issue of single point failure. The model parameters uploaded by the devices are taken as the transactions, and will be recorded in the candidate block after being verified by the associated miner. After all miners reached a consensus, devices can obtain the latest global model by aggregating the local model updates contained in the newly generated block downloaded from the associated miner. Generally, the global model is aggregated anywhere as long as the latest block can be obtained. In addition, the data reward for devices will be issued by the associated miner according to the size of the device data sample; the mining reward for miners will be issued by the blockchain network according to the total volume of data used by the connected devices. With the reasonable incentives,  the blockchain-based FL can be positively driven for efficient training.

\par Due to the decentralized architecture of BlockFL, the malfunction of each miner only distorts the global model of its own devices instead of paralyzing the entire system. Moreover, such distortion can be recovered by interaction with   other regular miners or federating with other devices associated with regular miners. Besides, by optimizing the block generation rate, the time for BlockFL to complete the model training can be reduced and the performance of the system can be improved.

\subsection{Related Work}
\subsubsection{Security}
Due to the capability of learning, analyzing and classifying, machine learning is used to monitor behaviors and
to detect the malicious attack for the blockchain security. S. Dey et al. in  \cite{chapter6RelatedWork1} combine machine learning and   game theory to solve the majority-attack problem. In this work,  the activities of attackers and participants in the network are monitored to  judge and classify  both participants' motivation and the service value in transactions, and then detect network anomalies. Therefore, the probability of   majority-attack will be reduced. H. Tang et al. in  \cite{chapter6RelatedWork2} introduce a deep learning-based algorithm to identify and classify malicious nodes by classifying behavior patterns in the network. The proposed algorithm can reduce the probability of the blockchain network being attacked by malicious nodes. Experimental results show that the proposed algorithm  is significantly more effective than the existing conventional methods. In order to detect anomalies (such as DDoS, double-spend and denial-of-service attacks) in electronic transactions of Bitcoin, S. Sayadi et al. in  \cite{chapter6RelatedWork3} propose an anomaly detection model based on machine learning.    This  work proposes two modes of machine learning approaches: $\left. {\rm{1}} \right)$ the One Class Support Vector Machines (OCSVM) algorithm to detect  outliers, $\left. {\rm{2}} \right)$  the K-Means algorithm in order to group outliers based on similiar type of anomalies.  Experimental results  show that the proposed model can accurately identify the types of attacks, and can provide the theoretical guidance to improve the security of the electronic trading system based on Bitcoin.
%机器学习非监督学习算法新增文献
P. Thai et al. in  \cite{chapter6RelatedWorAdd2} focus particularly on the anomaly detection  to  the Bitcoin transaction network, with the goal of detecting suspicious users and transactions. The data is first represented in two focal points: users and transactions. And then,  unsupervised learning methods including k-means,
%Mahalanobis distance based method,
and  SVM  are utilized to detect anomalies.
M. Shin et al. in   \cite{chapter6RelatedWorAdd3} propose a clustering method for bitcoin block and transaction data analysis, which defines the data that can be collected from the Bitcoin network, and the statistics of the blocks that can be extracted from the collected data. In addition, this paper performs a clustering experiment by applying Principal Component Analysis (PCA) to the extracted data, and also testes how to apply PCA to the clustering data.
\subsubsection{Performance improvement}
Machine learning is widely used for performance improvement of the blockchain systems. To meet the great demand of the blockchain, the solution  need to be found for the scalability problem. Nowadays, sharding method is found to resolve the scalability problem of the blockchain. A. Bugday et al. in  \cite{chapter6RelatedWork4} propose a method which uses adaptive machine learning model and Verifiable Random Functions together to assign nodes to achieve shards. As sharding method solves the scalability problem, the performance of the blockchain is improved. In order to optimize the system performance of the blockchain-based IoV, M. Liu et al. in \cite{chapter6RelatedWorkAdd1} use Deep Reinforcement Learning Technology (DLT) to select the consensus algorithm and the block generated nodes, as well as  adjust  the block size and the interval between block generations. The solution proposed can be applied to the dynamic IoV scenario, and it can maximie the system throughput without affecting the system's decentralization, latency and security.
%非监督学习新增文献
W. Hao et al. in  \cite{chapter6RelatedWorAdd4} establish a trust-enhanced blockchain P2P topology (BlockP2P-EP) that considers the transmission rate and transmission reliability to improve the performance of the blockchain network in achieving fast and reliable broadcasting. BlockP2P-EP first uses K-means to cluster neighboring peer nodes. Then on top of the trust-enhanced blockchain topology, BlockP2P-EP executes the parallel spanning tree broadcasting algorithm to achieve fast data broadcasting among nodes in terms of intra- and inter-clusters.

\subsubsection{Application in IoT}
Machine learning is also useful to apply blockchain in various fields such as IoT, IoV and smart grid. In order to implement blockchain technology into IoT fields and achieve the condition-based management on the blockchain, T. Id in  \cite{chapter6RelatedWork6} applies blockchain and machine learning into the task of anomaly detections in the IoT. The authors transform the collaborative anomaly detection task of the blockchain into multi-task probabilisitc dictionary learning. Then, the statistical machine learning algorithms is used to solve major technical issues such as building and validation of consensus in the blockchain. To ensure the performance of data aggregation, data storage, and data processing  in IoT services, N. C. Luong in  \cite{chapter6RelatedWork9} use blockchain to support IoT services which is based on cognitive radio network. To help secondary users (IoT devices) to choose an optimal transaction transmission under intricate conditions,  the authors get an optimal transaction transmission policy for secondary users by  adopting a Double Deep-Q Network (DDQN) algorithms that can allow the secondary users to learn the optimal policy above. H. Yao et al. in  \cite{chapter4RelatedWork7} introduce blockchain and cloud computing into IoT  to offload computational task from the IIoT network itself. In addition, this paper models the resource interaction between cloud providers and miners as a Stackelberg game, while proposing a multiagent reinforcement learning algorithm to find the Nash equilibrium of the proposed game.

\subsubsection{Application in smart grid}
To protect the smart grid from suffering cyber attacks, M. A. Ferrag et al. in  \cite{chapter6RelatedWork7} propose a novel deep learning and blockchain-based energy framework. This framework consists of two schemes: a blockchain based scheme and a deep learning-based scheme. The blockchain-based scheme is used to facilitate the exchange of excess energy among neighboring nodes. And the deep learning-based scheme is used to detect attacks and fraudulent transactions to enhance the system reliability and security.
\subsubsection{Application in VANETs}
For the problem that the data collected by different entities in the vehicle social network (VSNs) usually contains very different attributes, M. Shen et al. in \cite{8919978} propose a privacy-preserving SVM classifier training scheme over vertically-partitioned datasets posessed by multiple data providers.  In addition, consortium blockchain and threshold homomorphic cryptosystem are used to establish a secure SVM classifier training platform without a trusted third-party. To improve the security and reduce the attack in the vehicular ad hoc networks (VANETs), C. Dai et al. in  \cite{chapter6RelatedWork8} propose an indirect reciprocity security framework. This framework tries to encourage the On Board Units (OBUs) to help each others to reduce  attacks. According to a designed social norm, the framework assigns a scalar reputation to each OBUs to evaluate their dangerous level to the VANET, and apply the blockchain technique to protect the reputation from being
tampered. Each OBU  under the indirect reciprocity principle takes actions to another OBU based on the reputation which is assigned by the framework. Therefore,  a selection strategy for OBUs based on Q-learning (one kind of the reinforcement learning) is proposed to solve the core problem of action selection. According to the strategy, the OBU can achieve the optimal action.  Therefore, the proposed framework can efficiently increase both the reputation and the utility of the each OBU, and improve the security of the VANETs.

\subsection{Summary and Existing Problems}
In the literature, a number of works  have shown that using machine learning in both data management and analysis can effectively monitor and classify the behavior of blockchain users, as well as  recognize malicious behavior, suspicious user and illegal activity in the system, therefore both reduce the possibility of attacks and optimize the performance of system. Comparing with the traditional method of optimization or game theory, machine learning can adjust its strategy according to the changing of environment based on the long-term reward maximizing. However, there are still two existing problems that should be further investigated.
\begin{enumerate}
  \item Machine learning is valuable to understand the blockchain process and behavior in order to optimize system throughput and security (i.e., the transaction processing speed and malicious attack recognition), how to use the ability of machine learning in management, analysis and prediction to provide an intelligent guideline for decision-making and mechanism design is still an open issue.
  \item The impact of convergence and accuracy of selected machine learning algorithms on the performance and security of blockchain system should be considered.
\end{enumerate}

\section{Cryptography in Blockchain}

Cryptography \cite{chapter7Introduction8} is a method for confidential communication based on information transformation in a prescribed way, which is to ensure the confidentiality, integrity, authentication and non-repudiation. To improve information security, cryptography has been  widely used in security identification, such as access control and privacy protection. Symmetric encryption and asymmetric encryption are the mainstream cryptographic techniques to prevent information from being eavesdropped. Symmetric encryption use the same key for encryption and decryption, such as Data Encryption Standard (DES) \cite{chapter7Introduction2}, Advanced Encryption Standard (AES) \cite{chapter7Introduction1},  and International Data Encryption Algorithm (IDEA). Asymmetric encryption uses public and private keys to encrypt and decrypt, such as Rivest-Shamir-Adleman (RSA) \cite{chapter7Introduction5}\cite{chapter7Introduction6}, Elliptic-curve cryptography (ECC) \cite{chapter7Introduction3}, and Elgamal \cite{chapter7Introduction4}. Symmetric encryptionis fast and simple, but as the number of users increases, it will face the risk of key management and key leakage \cite{chapter7Introduction7}. In contrast, asymmetric encryption has high security but low efficiency \cite{AsymmetricEncryption}.

\subsection{Model Briefs}

Cryptography is the basic theory in blockchain, in which hashing, asymmetric encryption, timestamp, Merkle tree, etc. have been widely used to guarantee the security of transactions and the privacy of user information.
The basic structure of blockchain is chain-of-blocks, the chain can be seemed as the relationship of blocks, and the block is an encapsulated data structure to cache data/transactions. In order to maintain the blockchain system, blockchain users perform consensus mechanism to generate blocks continuously, and hash algorithm in cryptography is carried out for transaction integrity, proof of consensus, writing new block in chain, etc. Besides, a binary Merkle tree is used for data structure which can confirm the existence and integrity of the transactions quickly.
\begin{figure}[h]
  %\centering
  \includegraphics[width=3.5in]{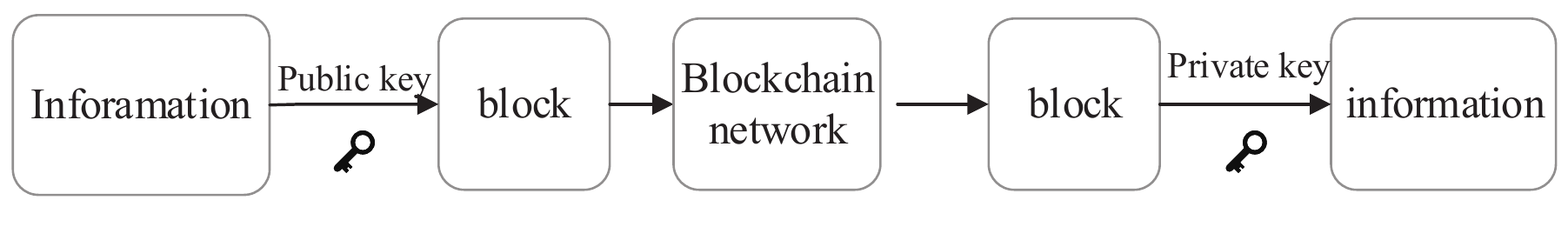}
  % where an .eps filename suffix will be assumed under latex,
  % and a .pdf suffix will be assumed for pdflatex; or what has been declared
  % via \DeclareGraphicsExtensions.
  \caption{ The Cryptography used in Blockchain.   }
  \label{fig-7-1}
\end{figure}
Fig. 17 shows the application of asymmetric encryption in the blockchain. with asymmetric encryption, blockchain users can use the public key to encrypt information in transaction to ensure the security. Meanwhile, the private key is used to sign the transaction digitally by any blockchain user, and the others can use the corresponding public key to   verify and to prevent in a distributed manner.  For the privacy issue caused by data stored in the blockchain,  some scheme can be adopted to provide anonymity decoupling users' information from published transactions, involving of
public key encryption scheme with keyword searchable, anonymous digital certificate publishing scheme, and zero-knowledge proof.
For data processing security in the blockchain, homomorphic encryption technology enables users to encrypt the transaction data using the corresponding encryption algorithm before submitting the transaction data to the block chain network. The data exists in ciphertext, which will not reveal any privacy information of users even if it is obtained by the attackers. Meanwhile, the ciphertext operation result is consistent with the plaintext operation result.
\begin{figure}[H]
  %\centering
  \includegraphics[width=3.5in]{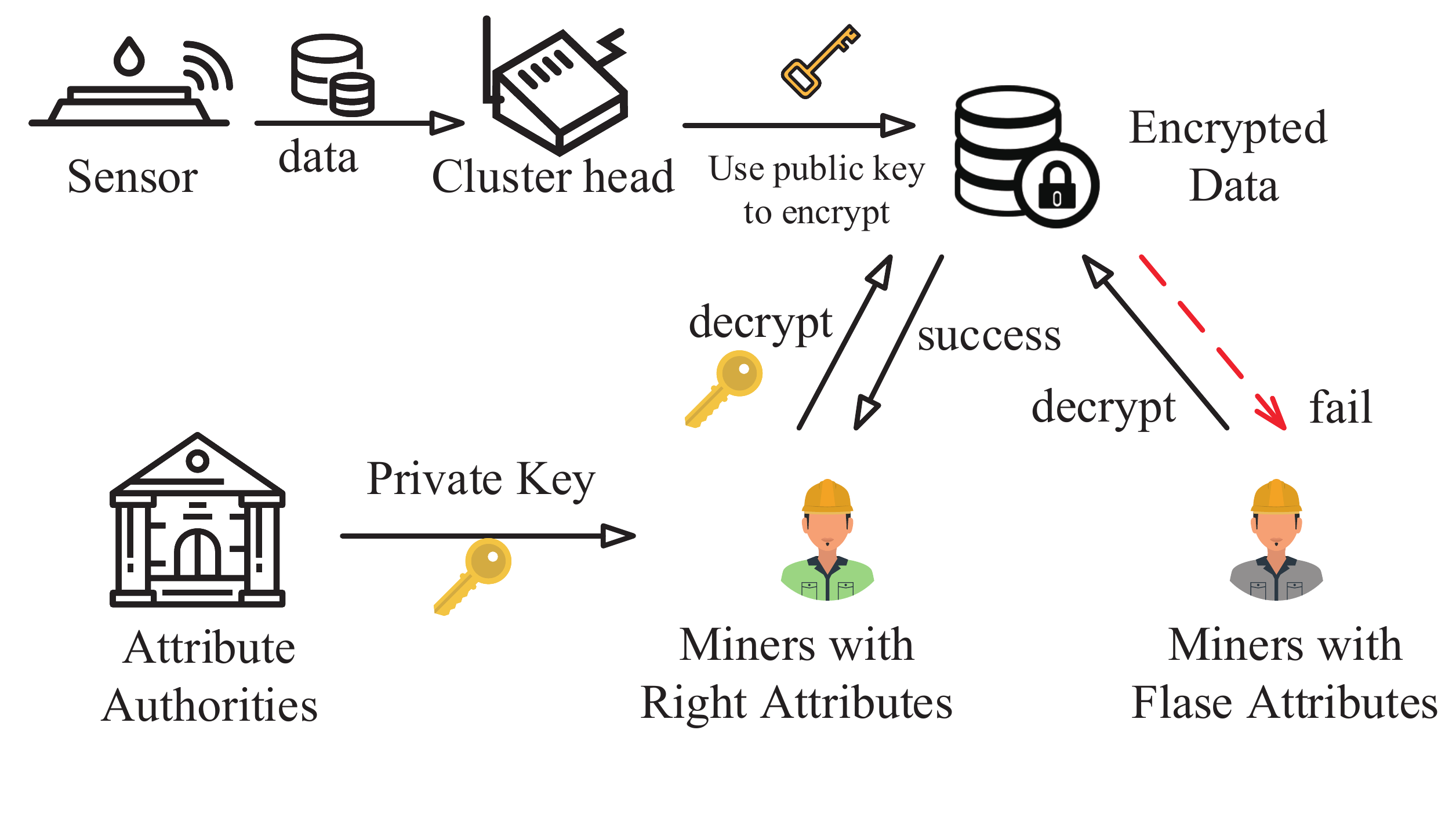}
  % where an .eps filename suffix will be assumed under latex,
  % and a .pdf suffix will be assumed for pdflatex; or what has been declared
  % via \DeclareGraphicsExtensions.
  \caption{Cryptography theory for access control and authentication in blockchain systems  }
  \label{fig-7-2}
\end{figure}
In the meantime, we can apply cryptography theory for access control and authentication in the blockchain systems. As shown in Fig. 18, by controlling the right of data reading and user access, only the users who have the right attributes can decrypted the encrypted data   and access the blockchain systems, while those users who do not have the right attribute cannot decypt data successfully. Thesefore,  the security of storing data and information in the blockchain can be guaranteed by using crytography.

\subsection{Case Study}
\par In this section,we use the previous work \cite{chapter7CaseStudy} to introduce the application of cryptography in blockchain. Due to the attributes of blockchain, many researchers try to apply blockchain technology to vehicular networks to improve its security. However, possible privacy leakage issues will reduce the enthusiasm of vehicles to participate in ad dissemination. In order to solve the vehicle privacy problem of the blockchain-based vehicular networks, the authors use Zero-knowledge proof of knowledge  (ZKPoK) to propose a new blockchain-based ad dissemination framework
which is shown in Fig. 19.
\par To ensure the validation of ad dissemination and improve the security of the vehicular networks, this work designs a concrete, fair and anonymous scheme under the proposed framework. In order to ensure fairness,  Merkle hash trees and smart contracts are used to implement the proof of ad reception to mitigate ``free-riding'' attacks. In addition,the proposed scheme can protect vehicles' privacy in terms of anonymity and conditional unlinkability based on zero-knowledge proof techniques. In addition, we briefly explain the five-stage process involved in the privacy protection blockchain architecture using Merkle hash tree and ZKPoK.
%the Merkle hash tree and smart contract are used, and can ensure anonymity and conditional linkability of vehicles issues by using ZKPoK for vehicle privacy. Accordingly, only vehicles that actually receive or forward the ad can receive the corresponding reward to avoid colluding with each other to falsify evidence to achieve "free-riding" attacks. Next, we briefly explain the five-stage process involved in the privacy protection blockchain architecture using Merkle hash tree and ZKPoK.
%\par Accordingly, the security and privacy issues in the current ad dissemination solution can be solved. Using the Merkle hash tree and smart contract, it ensures that vehicles can honestly participate in ad dissemination and mitigate "free-riding" attacks. In addition, the ZKPoK realizes the anonymity and conditional linkability of vehicles, thus realizing the privacy protection of vehicles.
%ABE algorithm contains five protocols. Now we will briefly explain the process of the five protocols used in the privacy-preserving blockchain architecture:
\begin{figure}[h]
  %\centering
  \includegraphics[width=3.3in]{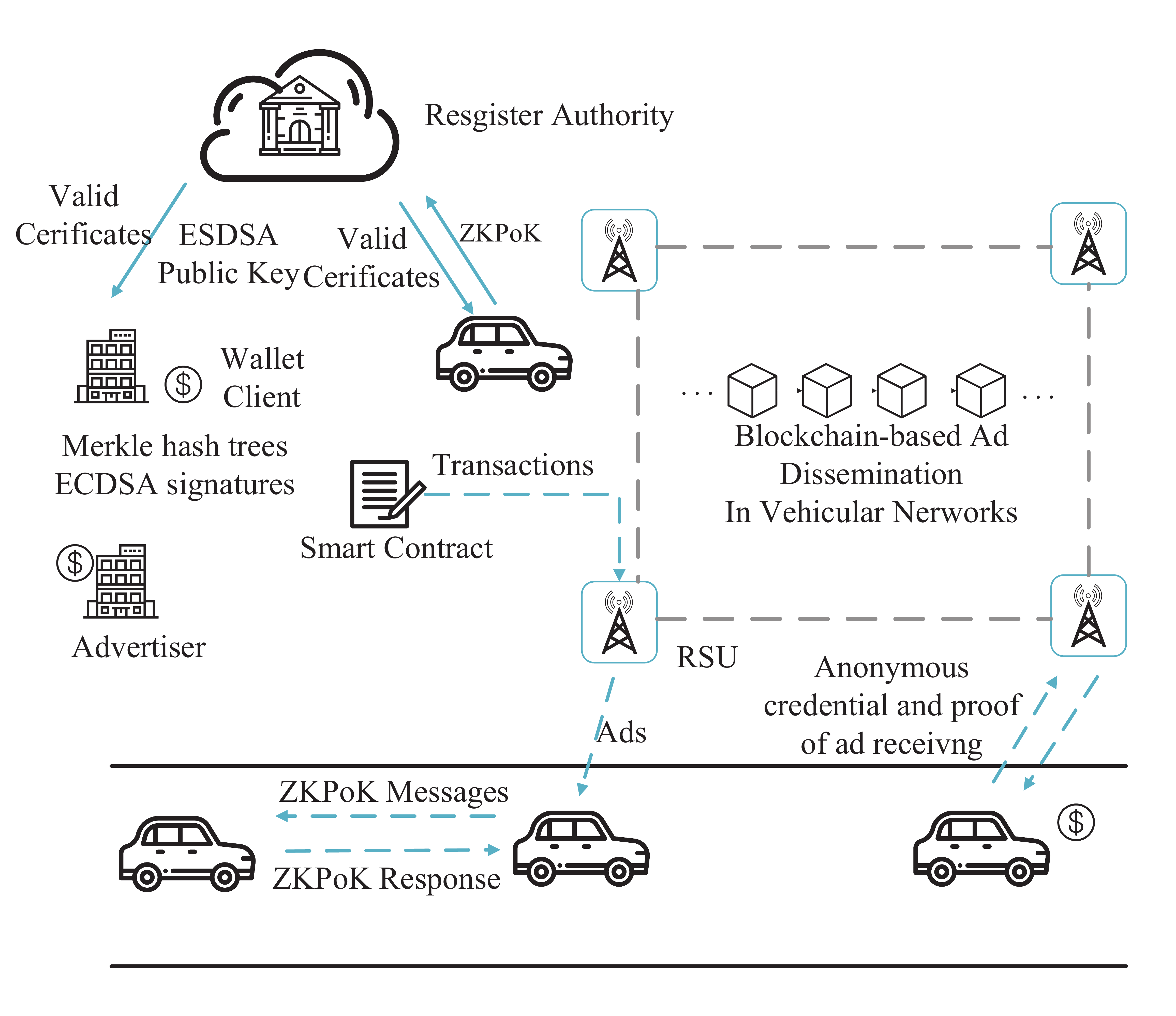}
  % where an .eps filename suffix will be assumed under latex,
  % and a .pdf suffix will be assumed for pdflatex; or what has been declared
  % via \DeclareGraphicsExtensions.
  \caption{ The blockchain-based AD Dissemination framework \cite{chapter7CaseStudy}}
  \label{fig_7-case_study}
\end{figure}
\begin{itemize}
  \item \textbf{System Setup:} Register authority initializes the system and generates public and private keys based on predefined security parameters.
  \item \textbf{Registration:} Advertisers and vehicles run sub-protocols to obtain valid certificates or keys individually. In the sub-protocol for advertisers, the registration authority adds registration information based on the ECDSA public key sent by the advertiser. In the sub-protocol for vehicle operation, register authority verifies the vehicle's identity information according to the ZKPoK sent by the vehicle.
        \item \textbf{Ad Publication: } Advertisers with valid certificates package ad information as a transaction and submit it to smart contracts, where the ad information is stored as the form of Merkle hash trees. Advertisers also generate ECDSA signatures so that participants can verify the validity of the transaction. Then, they select the suitable roadside unit to broadcast the ad.
  \item \textbf{Ad Dissemination:}  The selected Road Side Unit (RSU) broadcasts ad to nearby vehicles, or vehicles forward advertising to other vehicles. In addition, both the forwarding vehicle and the receiving vehicle will use ZKPoK to generate anonymous credentials to prove that the forwarding process  is actually took place.
  \item \textbf{Reward Payment:}. At this stage, RSU checks the validity of the anonymous credential and proof-of-ad-receiving submitted by the forwarding vehicle. If the verification is passed, the valid advertisement receiver and sender will receive corresponding rewards.
\end{itemize}

\subsection{Related Work}

\subsubsection{Security}
%密码学在安全中的应用新增文献
M. Noel et al. in  \cite{chapter7RelatedWorAdd1} present basic analysis and the background understanding of Stateful Hash-based Signature Schemes, particularly the Lamport One-Time Signature Scheme, Winternitz One-Time Signature Scheme, and the Merkle Signature Scheme. Moreover, the three hash-based digital signatures are compared in terms of key generation, signature generation, verification, and security levels.
K. Chalkias et al. in   \cite{chapter7RelatedWorAdd2} propose BPQS, a scalable post-quantum (PQ) digital signature scheme best suited for blockchain and distributed ledger technologies (DLTs). It can leverage application-specific chain/graph structures to reduce the cost of  key generation, signing and validation, as well as decreasing the  size of signature. Besides, an open source implementation of the scheme is provided in this paper. Furthermore, compared with other benchmark schemes (e.g., SHA256 and SHA384), BPQS is superior to existing hashing algorithms when reusing keys for a reasonable number of signatures. The reason is that it support a fallback mechanism to allow for an almost unlimited number of signatures if needed.
M. Sato et al. in  \cite{chapter7RelatedWorAdd3} propose a scheme to extend the validity of past blocks when the underlying cryptographic algorithms (hash functions and digital signatures) are destroyed. This scheme can effectively avoid the hard-fork of the original blockchain when the hash function is compromised, and provide the smooth-fork when the digital signature scheme  is compromised.

\subsubsection{Privacy}

Cryptography is widely used to protect the privacy of users in the blockchain. G. Micaliey al. in  \cite{chapter7RelatedWork1} utilize the partial homomorphic encryption to enhance data privacy, and resist  both collision attack and prime attack in the blockchain. In order to protect the privacy of blockchain users, P. Zhong et al in  \cite{chapter7RelatedWork2} design a Privacy-Protected blockchain system to protect the privacy of users in the blockchain. This system encrypts all data uploaded by blockchain users within the agreed time. During this period, the user's personal data will not leak unless he voluntarily publish these data. When the agreed time is reached, the system publishes the decryption key and verifies the user's behavior to avoid fraudulent behavior by malicious users. Y. Rahulamathavan et al. in  \cite{chapter7RelatedWork3}   apply attribute-based encryption to control the permission of data access and usage in the blockchain system to achieve privacy protection. M. Zhang et al. in  \cite{chapter7RelatedWork4}  propose an ID-based encryption scheme to improve the data privacy particularly for the non-transaction applications in the blockchain. This scheme encrypts the plaintext into the ciphertext to hide the information for preventing disguise and passive attacks. One innovation of this sheme is that complex certificate management and issuance in traditional PKI systems can be avoid without using advance technologies such as zero-knowledge proof.

\subsubsection{Zero-knowledge proof}

Zero-knowledge proof (ZKP) provides the ability to prove secrets without directly disclosing them. It guarantees that the proof will not reveal more information about private input, nor can it infer from the calculations. A large number of research scholars have used ZKP to solve blockchain privacy issues. M. Harikrishnan et al. in  \cite{chapter7RelatedWork6} study the problem of confidentiality of data in the blockchain network and design a new interactive ZKP encryption technology-ZKSTARK, which is achieved by selecting two indistinguishable hash functions in the blockchain system and integrating them into the ZKP protocol. M. H. MurtazaIn et al in \cite{chapter7RelatedWork7} introduce a simple non-interactive ZKP scheme for blockchain-based electronic voting systems, namely zkSNARKs. This work  uses digital signatures for  message authentication,   cryptographic hash functions for getting a message digest, and  ZKP for obtaining unlinkability, to achieve the security and confidentiality of the electronic voting system. On this basis, D. Ding et al. in  \cite{chapter7RelatedWork8} propose a blockchain privacy protection scheme using accounts and multi-asset model, where the zkSNARKs algorithm is use  to generate and to verify the zero-knowledge proof. This scheme is faster than ZKSTARK, but it requires trusted settings, unable to resist quantum attacks, and is less secure than ZKSTARK. Y. Tsai et al. in \cite{chapter7RelatedWork9} introduce an improved non-interactive zero-knowledge range proof scheme based on the predecessors. This work  use the Fujisaki-Okamoto commitment, non-interactive zero-knowledge and computational bindingness through proof of knowledge in the cyclic group with secret order technique (CBPKCGSO), to achieve better security and efficiency.

\subsubsection{Verification}
A. S. Sani  et al. in  \cite{8885084}  introduce a new high-performance and scalable blockchain network. This blockchain network uses Time-based Zero-Knowledge Proof of Knowledge (T-ZKPK) to perform identity verification and the establishment of key protection transactions, which can enhance the security and privacy of IIoT.
%ADD 1-2 WORK
%新增验证相关文献
S. Zhu et al. in  \cite{chapter7RelatedWorAdd4}  propose a novel hybrid blockchain crowdsourcing platform to achieve decentralization and privacy preservation. On this platform, a hybrid blockchain structure, dual-ledgers and dual-consensus algorithms are integrated to ensure secured communication between requesters and workers. In addition, the smart contract and zero-knowledge proof are deployed to achieve permission control and privacy protection.
Z. Wan et al. in   \cite{chapter7RelatedWorAdd5} design a zero-knowledge SNARK scheme for authenticated data by effectively combining zk-SNARK technology with digital signature, called zk-DASNARK. Based on this, zk-AuthFeed (a zero-knowledge authenticated data feed scheme) is designed to achieve data privacy and authenticity of smart contracts, and it is implemented using libsnark and Ethereum to prove the efficiency of the proposed scheme.

\subsubsection{Supervisability}
To solve the nonsupervisability problem in the blockchain-based IoT e-commerce autonomous transaction management system, C. Liu et al. in  \cite{chapter7RelatedWork5} propose a new transaction settlement system called NormaChain. This work design  a three-layers sharding blockchain network  and an innovative decentralized public key searchable encryption scheme (decentralized public key encryption with keyword search (PEKS) scheme)  to resist chosen ciphertext attacks, cryptanalysis, and collusion. H. Kang et al. in  \cite{chapter7RelatedWork11} propose  a blockchain network called FabZK. By only storing the encrypted data of each transaction, and anonymizing the transaction relationship between participants  at the same time, this blockchain network FabZK hides the transaction details in the shared ledger. In addition, it achieves both privacy and auditability by supporting verifiable Pedersen commitments and constructing zero-knowledge proofs. Evaluation
result shows that this FabZK offers strong privacy-preserving capabilities, while delivering reasonable performance for the applications developed based on its framework.
% \usepackage{multirow}
\begin{table*}[]
  \caption{Reference classification on methodology perspective}
  \label{Tab2}
  \centering
  \begin{tabular}{|c|c|c|c|c|}
    \hline
    \multicolumn{2}{|l|}{\diagbox{Method}{Research category}} & Theoretical model               & Network service                               & Application                                                                                                                                                        \\ \hline
    \multirow{4}{*}{Stochastic}                               & \begin{tabular}[c]{@{}c@{}}Markov \\ process\end{tabular}  & -                                             & -                                                                                                           & \begin{tabular}[c]{@{}c@{}}performance and security \\ analysis in IoT {\cite{chapter3CaseStudy}}\end{tabular}                       \\ \cline{2-5}
                                                              & \begin{tabular}[c]{@{}c@{}}Poisson \\ process\end{tabular}  & \begin{tabular}[c]{@{}c@{}}mining difficulty \\ control {\cite{chapter3RelatedWork1}\cite{chapter3RelatedWork2}}\end{tabular}                & -                                                                                                           & \begin{tabular}[c]{@{}c@{}}node deployment in IoT {\cite{chapter3RelatedWork6}},\\ node deployment in MEC {\cite{chapter3RelatedWork7}}\end{tabular}                       \\ \cline{2-5}
                                                              & \begin{tabular}[c]{@{}c@{}}Stochastic \\ geometry\end{tabular}  & -                                             & -                                                                                                           & node deployment in IoT {\cite{chapter3RelatedWork5}} \\ \cline{2-5}
                                                              & Others                          & \begin{tabular}[c]{@{}c@{}}modeling and performance \\ analysis {\cite{chapter3RelatedWork3}\cite{chapter3RelatedWork4}}\end{tabular}                & -                                                                                                           & -                                                    \\ \hline
    \multirow{5}{*}{Game}                                     & \begin{tabular}[c]{@{}c@{}}Stackelberg \\ game\end{tabular}  & -                                             & \begin{tabular}[c]{@{}c@{}}incentive mechanismin  IoT {\cite{chapter4CaseStudy}}, \\security strategy in network {\cite{chapter4RelatedWork5}}, \\resource  management in MEC {\cite{chapter4RelatedWork7}}\end{tabular}                                                                              & -                                                    \\ \cline{2-5}
                                                              & \begin{tabular}[c]{@{}c@{}}Evolutionary \\ game\end{tabular}  & \begin{tabular}[c]{@{}c@{}}mining pool \\ management {\cite{chapter4RelatedWork1}},\\ mining pool \\ security strategy {\cite{chapter4RelatedWork6}}\end{tabular}                & -                                                                                                           & -                                                    \\ \cline{2-5}
                                                              & \begin{tabular}[c]{@{}c@{}}Iterative \\ game\end{tabular}  & \begin{tabular}[c]{@{}c@{}}mining pool \\ management {\cite{chapter4RelatedWork3}\cite{chapter4RelatedWork4}}\end{tabular}                & -                                                                                                           & -                                                    \\ \cline{2-5}
                                                              & Auction                         & -                                             & \begin{tabular}[c]{@{}c@{}}resource management \\ in MEC {\cite{chapter4RelatedWork8}\cite{chapter4RelatedWork9}}\end{tabular}                                                                              & -                                                    \\ \cline{2-5}
                                                              & Others                          & \begin{tabular}[c]{@{}c@{}}mining pool \\ management {\cite{chapter4RelatedWork2}}\end{tabular}                & \begin{tabular}[c]{@{}c@{}}security \\ in edge networks {\cite{chapter4RelatedWork10}}\end{tabular}                                                                              & -                                                    \\ \hline
    \multirow{5}{*}{Optimization}                             & \begin{tabular}[c]{@{}c@{}}Convex \\ optimization\end{tabular}  & -                                             & \begin{tabular}[c]{@{}c@{}}security  in D2D \\ communication {\cite{chapter5CaseStudy}}\end{tabular}                                                                              & -                                                    \\ \cline{2-5}
                                                              & \begin{tabular}[c]{@{}c@{}}Geometric \\ programming\end{tabular} & -                                             & \begin{tabular}[c]{@{}c@{}}resource allocation \\ in IoT {\cite{chapter5RelatedWork4}}\end{tabular}                                                                             & -                                                    \\ \cline{2-5}
                                                              & \begin{tabular}[c]{@{}c@{}}Stochastic \\ programming\end{tabular} & -                                             & \begin{tabular}[c]{@{}c@{}}optimal algorithm and \\ strategy design in mobile \\ edge network {\cite{chapter5RelatedWorAdd2}}\end{tabular}                                                                             & -                                                    \\ \cline{2-5}
                                                              & \begin{tabular}[c]{@{}c@{}}Lyapunov \\ Optimization\end{tabular} & DDoS attack  avoidance {\cite{chapter5RelatedWork2}}     & \begin{tabular}[c]{@{}c@{}}resource allocation in \\ mobile device cloud {\cite{chapter5RelatedWorAdd1}}\end{tabular}                                                                             & -                                                    \\ \cline{2-5}
                                                              & Others      &      \begin{tabular}[c]{@{}c@{}}analytical framework modeling   \\  for PBFT {\cite{9013778}}\end{tabular}                                          & \begin{tabular}[c]{@{}c@{}}security in IoV {\cite{chapter5RelatedWork1}}, \\ resource allocation in  MEC {\cite{chapter5RelatedWork3}}, \\ optimal algorithm  and strategy design in \\ payment channel network {\cite{chapter5RelatedWork5}}\end{tabular}                                                                             & \begin{tabular}[c]{@{}c@{}}optimal algorithm and \\ strategy design in electric \\ taxi charging scenarios {\cite{chapter5RelatedWork6}}\end{tabular}                      \\ \hline
    \multirow{7}{*}{\begin{tabular}[c]{@{}c@{}}Machine \\ Learning\end{tabular}}          & \begin{tabular}[c]{@{}c@{}}Supervised \\ learning\end{tabular} & majority-attack  avoidance {\cite{chapter6RelatedWork1}} & -                                                                                                           & -                                                    \\ \cline{2-5}
                                                              & \begin{tabular}[c]{@{}c@{}}Unsupervised \\ learning\end{tabular} & \begin{tabular}[c]{@{}c@{}}performance \\ optimization {\cite{chapter6RelatedWorAdd4}}\end{tabular}               & security in Bitcoin {\cite{chapter6RelatedWork3}\cite{chapter6RelatedWorAdd2}\cite{chapter6RelatedWorAdd3}} & -                                                    \\ \cline{2-5}
                                                              & \begin{tabular}[c]{@{}c@{}}Federated \\ learning\end{tabular} & -                                             & \begin{tabular}[c]{@{}c@{}}privacy and security in \\ centralized machine \\ learning {\cite{chapter6CaseStudy}}\end{tabular}                                                                             & -                                                    \\ \cline{2-5}
                                                              & \begin{tabular}[c]{@{}c@{}}Deep \\ learning\end{tabular} & \begin{tabular}[c]{@{}c@{}}identify malicious \\ nodes {\cite{chapter6RelatedWork2}}\end{tabular}               & -                                                                                                           & \begin{tabular}[c]{@{}c@{}}application in IoT {\cite{chapter6RelatedWork9}}, \\ application in smart grid {\cite{chapter6RelatedWork7}}\end{tabular}                      \\ \cline{2-5}
                                                              & \begin{tabular}[c]{@{}c@{}}Reinforcement \\ learning\end{tabular} & -                                             & \begin{tabular}[c]{@{}c@{}}resource management \\ in IoT {\cite{chapter4RelatedWork7}}\end{tabular}                                                                             & -                                                    \\ \cline{2-5}
                                                              & \begin{tabular}[c]{@{}c@{}}Deep \\ reinforcement \\ learning\end{tabular} & -                                             & security in IoV {\cite{chapter6RelatedWork8}}                                                               & \begin{tabular}[c]{@{}c@{}}performance optimization\\ in IoV {\cite{chapter6RelatedWorkAdd1}}\end{tabular}                      \\ \cline{2-5}
                                                              & Others                          & -                                             & \begin{tabular}[c]{@{}c@{}}sharding management {\cite{chapter6RelatedWork4}},\\ collaborative anomaly \\ detection in IoT {\cite{chapter6RelatedWork6}}\end{tabular}                                                                             & \begin{tabular}[c]{@{}c@{}}SVM traning platform       \\ for VSNs  {\cite{8919978}}\end{tabular}                        \\ \hline
    \multirow{4}{*}{Cryptography}                             & \begin{tabular}[c]{@{}c@{}}Asymmetric \\ encryption\end{tabular} & -                                             & \begin{tabular}[c]{@{}c@{}}security issues {\cite{chapter7RelatedWorAdd1}\cite{chapter7RelatedWorAdd2}}, \\ user privacy  {\cite{chapter7RelatedWork2}}\end{tabular}                                                                             & -                                                    \\ \cline{2-5}
                                                              & \begin{tabular}[c]{@{}c@{}}Partial \\ homomorphic \\ encryption\end{tabular} & -                                             & user privacy   {\cite{chapter7RelatedWork1}}                                                           & -                                                    \\ \cline{2-5}
                                                              & ZKP                             & -                                             & \begin{tabular}[c]{@{}c@{}}privacy {\cite{chapter7CaseStudy}\cite{chapter7RelatedWork6}\cite{chapter7RelatedWork7}}\\ \cite{chapter7RelatedWork8}\cite{chapter7RelatedWork9}\cite{chapter7RelatedWork11}, \\ identity   verification{\cite{8885084}\cite{chapter7RelatedWorAdd4}\cite{chapter7RelatedWorAdd5}}\end{tabular}                                                                             & -                                                    \\ \cline{2-5}
                                                              & Others                          & -                                             & \begin{tabular}[c]{@{}c@{}}security {\cite{chapter7RelatedWorAdd3}\cite{chapter7RelatedWork5}}, \\ privacy   {\cite{chapter7RelatedWork3}\cite{chapter7RelatedWork4}}\end{tabular}                                                                             & -                                                    \\ \hline
  \end{tabular}
\end{table*}

\subsection{Summary and Existing Problem}
Cryptography is the key element for confidentiality, integrity, authentication and non-repudiation for security and privacy in blockchain using the typical scheme like digital signature, asymmetric encryption, hashing and etc. These methods focus on security and privacy protection in an external manner, which are to design a powerful wall around the blockchain system to deny the attacks from malicious users. However, the system would be vulnerable if any bugs found by the attacker to break. Therefore, how to empower the security and privacy ability of blockchain system in a systematic manner providing inherent safety is still an open issue. Moreover, the mechanism and protocol should be designed to optimize the performance as well as security simultaneously.

\section{Open Issue Discussions}
This paper discussed the theoretical model research of blockchain basic knowledge, the design of network services based on blockchain mechanisms and algorithms, and the deployment of blockchain-based applications in practical systems from a methodological perspective, and the reference classification is shown in Table II. In the meantime, some remaining problems in terms of technical,
commercial and political views are still needed to be discussed as the open issues.

\subsection{Technology Issue}

%从通信传输来看,通信是区块链信息的承载与通道,要获得区块链带来的优势,势必会带来额外的通信开销,这部分讨论区块链运行导致的通信开销原因（交易发布、共识交互、账本更新等）以及影响（区块链角度：对扩展性、去中心化挑战、安全性不可能三角的影响；通信网络角度：吞吐量、时延、QoS、 公平性等）。

%从资源消耗来看,区块链经过多年的发展改进,在算力消耗方面已大有改进,但大规模应用中,以算力为代表的资源消耗依然极具挑战,特别是5G/B5G的能耗问题已经是一个重大负担,在引入区块链必然会加剧这一问题,必须得到重视。然而,虽然已提出相当多低算力开销的区块链算法,比如PoS、PBFT、DAG,但存在扩展性、安全性、去中心化等问题,并且现在大量采用的PBFT、DAG等算法需要大量投票广播交互,可以视为一种以通信开销换取算力消耗的折中,对区块链应用的承载网络依然不适用。

%未来研究思考如何精确评估计算区块链带来的通信开销及其影响；如何平衡区块链优势及其代价；如何设计适用于以5G/B5G为代表的新型通信网络的区块链算法协议,关注区块链移动性、轻量化、大规模流量、异构等新特性,利用现有基础设施和协议机制携带区块链信息,快速、安全、高效传输。另外,绿色区块链也是另一个值得重点关注的问题。
%????
From the resource consumption perspective,   many blockchain protocols, which consumes less power than PoW, have been proposed, such as PoS, IOTA and so on. However, they struggle to be widely used in wireless networks because of  their scalability. As to another type of blockchain protocols which vote instead of calculating, such as PBFT, Hashgraph and so on, they are still not widely applicable to wireless networks due to their communication complexity. From the communication perspective, caused by blockchain's transaction release, consensus interaction, ledger updating and so on, additional communication overhead is bound to be incurred along with the improvement of data security brought by blockchain. There have to be a tradeoff between the communication performance and the security performance of the blockchain network. In conclusion, essential questions to be technologically addressed include: $\left. {\rm{1}} \right)$ how to design a blockchain protocol with high scalability and appropriate power consumption, communication complexity while ensuring its security, $\left. {\rm{2}} \right)$ how to compromise between blockchain's performance and network's performance.

\subsection{Commercial Issue}

%现今区块链还缺乏杀手级应用,商业模式不明晰。虽然很多研究关注区块链在各行业的应用,但至今缺乏成熟商业应用,1）基础设施可用性方面,交易性能不足（TPS太低）、交易费用太高,各方面性能尚未满足基本的商用门槛；2）应用开发环境方面：界面不友好,开发壁垒高、太多现实商业场景无法实现应用,类似云服务从IAAS 发展到PAAS、SAAS 的过程中遇到的问题；3）用户群体方面：受众群体狭窄,与现实生活结合不够,技术还有待成熟。这些都影响了区块链技术在商业应用的大规模实现。

%私有链/联盟链来讲,杀手级的商业应用应对数据和流程的真实性和可靠性有需求,而且这真实性和可靠性能够创造很大价值的地方,如金融、溯源等行业。同时,区块链只能解决数字世界的信任问题,上链前数据的真实性需要得到保证,因此还需要有物联网等可以实时采集数据并保证数据真实性的条件。

%公有链来讲,更多的是对整个体系的改造,适合用在缺乏有效激励体系或可靠分配机制的系统。最先落地的领域会在那些已经形成了共识但缺乏激励机制、没有办法大规模落地的领域。

In addition to the technical matters, the lack of killer applications \cite{killerapplication} and business models also hinders the large-scale application of blockchain. For private or consortium blockchains, only  the untamperability of the data on chain can be guaranteed.  While   the authenticity of the data off chain requires the cooperation of other technologies form Internet of Things. In terms of public chain, it is more about the transformation of the whole system, which is suitable for the system without effective incentive system or reliable allocation mechanism. The first areas to to be implemented  will be those   that have formed a consensus but lack incentives and cannot be implemented on a large scale.
%For public blockchain, we suggest that the first breakthrough will be in the distributed systems, which are lack of reliable incentive mechanisms.

\subsection{Policy Issue}

%阻碍区块链发展和大规模商用的还有政策法规监管和标准统一问题。技术纷标准繁林立的情况下,需要从政府机构、行业组织和公司平台方面开放标准、程序代码以及跨平台协作,加速大规模区块链技术采用。政策法规和标准制定需要考虑区块链技术的发展和局限性,同时,区块链在解决方案选取和技术路线等发面也必须考虑政策法规和标准的限制。

The lack of policies, regulations, standardization and other related policy issues are  hindering the mass commercialization of blockchain.   In the formulation of them, the development and limitations of blockchain should be taken into consideration. At the mean time, the selection of solutions in blockchain must also comply with the restrictions of policies, regulations and standards.

\subsection{Communication Issue}
Blockchain is built in a P2P network, which means that a large amount of communication overhead will be added in terms of network traffic and system processing capacity. Because current applications are highly dynamic, data sources frequently change. This indicates that the node may have to send a large number of update transactions, which will further increase the communication overhead. On the other hand, blockchain deployment in wireless is foreseeable in the near future \cite{WireslessResouce}. In this case, blockchain services rely on wireless communication networks to reach consensus. During the consensus process, blockchain nodes are connected through wireless channels. However, due to factors such as wireless channel fading and unauthorized malicious interference, the wireless connection among blockchain nodes may be attacked, and the uplink or downlink transmission may fail, thereby reducing the probability of successful transactions.
Therefore, in the wireless blockchain system, a framework is needed to measure the communication overhead and communication quality in the communication process.

\section{Conclusion}
Blockchain is an emerging technology that is considered as one of the key enablers of 5G networks due to its unique features of decentralization, scalability, security and the corresponding characteristics. In this article, we presented a comprehensive survey focusing on the current most advanced achievements in exploring the intrinsic nature of blockchain from a methodological perspective. Based on the state of art literatures, we outlined the theoretical model research for blockchain fundamentals understanding, the network service design for blockchain-based mechanisms and algorithms, as well as the application of blockchain for Internet of Things and etc. We first introduced the working principles, research activities, and challenges of blockchain, as well as illustrated the roadmap involving the classic methodology with typical blockchain use cases and topics. Subsequently, we discussed the contribution of the methodology to the performance of blockchain systems, focusing on the role of  stochastic process, game theory, optimization, machine learning and cryptography in both the study of blockchain operation process and the design of blockchain protocol/algorithm. Finally, we pointed out several blockchain issues from technical, commercial, and political perspectives. Although the blockchain is still in its infancy, it is clear that blockchain will significantly improve the landscape and experience of future network services and applications.

% or


%\appendices

%\section{}

%\section*{Acknowledgment}


\ifCLASSOPTIONcaptionsoff
  \newpage
\fi



%\bibliographystyle{IEEEtran}
%\begin{thebibliography}{1}


%\end{thebibliography}
\bibliographystyle{IEEEtran}
\bibliography{references}
%\end{IEEEbiography}
%\newpage
\begin{IEEEbiographynophoto}{Bin Cao}
is an associate professor in the state key laboratory of network and switching technology at Beijing University of Posts and Telecommunications (BUPT). Before that, he was an associate professor at Chongqing University of Posts and Telecommunications. He received his Ph.D. degree (Honors) in communication and information systems from the National Key Laboratory of Science and Technology on Communications, University of Electronic Science and Technology of China in 2014. From April to December in 2012, he was an international visitor at the Institute for Infocomm Research (I2R), Singapore. He was a research fellow at the National University of Singapore from July 2015 to July 2016. He served as a guest editor for IEEE Sensors Journal and IEEE Transactions on Industrial Informatics, he also served as symposium cochair for IEEE ICNC 2018, blockchain workshop cochair for CyberC 2019, IEEE Blockchain 2020 special session and TPC member for numerous conferences. His research interests include blockchain system, internet of things and mobile edge computing.
\end{IEEEbiographynophoto}

%\newpage

\begin{IEEEbiographynophoto}{Zixin Wang}
received the M.E degree in information and communication engineering from Chongqing University of Posts and Telecommunications, Chongqing, China, in 2020. She currently is pursuing her Ph.D. degree  in
the State Key Laboratory of Networking and Switching Technology,
Beijing University of Posts and Telecommunications, Beijing, China. Her
research interests include blockchain and Internet of Things.
\end{IEEEbiographynophoto}

\begin{IEEEbiographynophoto}{Long Zhang}
received the M.E degree in information and communication engineering from Chongqing University of Posts and Telecommunications, Chongqing, China, in 2019. He currently is pursuing his Ph.D. degree at the National Key Laboratory of Science and Technology on Communications, University of Electronic Science and Technology of China, Chengdu, China. His research areas include next generation mobile networks and Internet of Things.
\end{IEEEbiographynophoto}

\newpage

\begin{IEEEbiographynophoto}{Daquan Feng}
received the Ph.D. degree in information engineering from the University of Electronic Science and Technology of China in 2015. He had been a visiting student with the School of Electrical and Computer Engineering, Georgia Institute of Technology, USA, from 2011 to 2014. After graduation, he was a research staff in the State Radio Monitoring Center, Beijing, China, and then a Postdoctoral Research Fellow in Singapore University of Technology and Design. He is now an Assistant Professor with the Shenzhen Key Laboratory of Digital Creative Technology, the Guangdong Province Engineering Laboratory for Digital Creative Technology, College of Electronics and Information Engineering, Shenzhen University, Shenzhen, China. His research interests include blockchain technology, URLLC communications, MEC, and massive IoT networks. He is an Associate Editor of IEEE COMMUNICATIONS LETTERS.
\end{IEEEbiographynophoto}

\begin{IEEEbiographynophoto}{Mugen Peng}
  received the Ph.D. degree in communication and information systems from the Beijing University of Posts
and Telecommunications (BUPT), Beijing, China, in 2005. Afterward, he
joined BUPT, where he has been a Full Professor with the School of
Information and Communication Engineering since 2012. In 2014, he
was an Academic Visiting Fellow with Princeton University, Princeton,
NJ, USA. He leads a Research Group focusing on wireless transmission
and networking technologies with the State Key Laboratory of Networking and Switching Technology, BUPT. He has authored/coauthored over
100 refereed IEEE journal papers and over 300 conference proceeding
papers. Dr. Peng was a recipient of the 2018 Heinrich Hertz Prize Paper
Award, the 2014 IEEE ComSoc AP Outstanding Young Researcher
Award, and the Best Paper Award in the JCN 2016 and IEEE WCNC
2015. He is on the Editorial/Associate Editorial Board of the IEEE Communications Magazine, the IEEE Internet of Things Journal,
and IEEE Access.
\end{IEEEbiographynophoto}

\begin{IEEEbiographynophoto}{Lei Zhang}
 is a Senior Lecturer (Associate Professor) at the University of Glasgow, U.K. He received his Ph.D. from the University of Sheffield, U.K. His research interests include wireless communication systems and networks, blockchain technology, radio access network slicing (RAN slicing), Internet of Things (IoT), multi-antenna signal processing, MIMO systems, etc. He has 19 patents granted/filed in more than 30 countries/regions including US/UK/EU/China/Japan etc. Dr Zhang has published 2 books and 100+ peer-reviewed papers. He received IEEE Communication Society TAOS TC Best Paper Award 2019. Dr. Zhang is a senior member of IEEE. He is a Technical Committee Chair of 5th International conference on UK-China Emerging Technologies (UCET) 2020. He was the Publication and Registration Chair of IEEE Sensor Array and Multichannel (SAM) 2018, Co-chair of Cyber-C Blockchain workshop 2019. He is an associate editor of IEEE Internet of Things (IoT) Journal, IEEE Wireless Communications Letters and Digital Communications and Networks.
\end{IEEEbiographynophoto}


%\vfill


\end{document}